\documentclass[journal]{IEEEtran}

\usepackage{multicol}
\usepackage{indentfirst}
\usepackage[pdftex]{graphicx}
\usepackage{amsthm}
\usepackage[fleqn]{amsmath}
\usepackage{amssymb}
\usepackage{amsfonts}
\usepackage{mathrsfs}
\usepackage{graphicx}
\usepackage{enumerate}
\usepackage{multirow}
\usepackage{booktabs}
\usepackage{enumitem}
\usepackage{enumerate}
\usepackage[noend]{algpseudocode}
\usepackage{algorithmicx,algorithm}
\usepackage{bm}

\newtheorem{Theorem}{Theorem}
\newtheorem{Definition}{Definition}

\newtheorem{Corollary}{Corollary}

\newtheorem{Remark}{Remark}

\newtheorem*{Square Root Law}{Square Root Law}
\newtheorem{Lemma}{Lemma}

\ifCLASSINFOpdf
\else
\fi

\begin{document}
\title{Bounds on $f$-Divergences between Distributions within Generalized Quasi-$\varepsilon$-Neighborhood}

\author{Xinchun~Yu,
  Shuangqing Wei, and Xiao-Ping Zhang 
\thanks{Xinchun Yu is with Zhejiang GongShang University, Hangzhou, 310018, China (e-mail: yxc@mail.zjgsu.edu.cn). Shuangqing Wei is with the Division of Electrical and Computer Engineering
School of Electrical Engineering and Computer Science,
Louisiana State University, Baton Rouge, LA 70803, USA (e-mail: swei@lsu.edu). Xiao-Ping Zhang is with Shenzhen Key Laboratory of Ubiquitous Data Enabling, Shenzhen International Graduate School, Tsinghua University, Shenzhen, 518055, China (e-mail: xpzhang@ieee.org).
 .

}
}
\maketitle

\begin{abstract}
This work establishes computable bounds between f-divergences for probability measures within a generalized quasi-$\varepsilon_{(M,m)}$-neighborhood framework. We make the following key contributions. (1) a unified characterization of local distributional proximity beyond structural constraints is provided, which encompasses discrete/continuous cases through parametric flexibility. (2) First-order differentiable $f$-divergence classification with Taylor-based inequalities is established, which generalizes $\chi^2$-divergence results to broader function classes. (3) 
We provide tighter reverse Pinsker's inequalities than existing ones, bridging asymptotic analysis and computable bounds. The proposed framework demonstrates particular efficacy in goodness-of-fit test asymptotics while maintaining computational tractability.

\end{abstract}

\begin{IEEEkeywords}
local information geometry, $f$-divergence, total variational distance, reverse Pinsker's inequality
\end{IEEEkeywords}
\section{Introduction}
The study of relationships between probability distribution measures has long been a central focus in probability theory, statistics, and information theory \cite{Csiszar0}-\cite{Sreekumar}. Establishing bounds between these measures proves particularly crucial for analyzing convergence rates in statistical methods \cite{Guntuboyina}\cite{Guntuboyina2} and exhibits strong connections to machine learning theory \cite{Reid}\cite{Reid2}. While existing literature predominantly addresses universal bounds for $f$-divergences without distributional constraints \cite{Csiszar2}-\cite{Sason}, the local behavior of these divergences holds significant importance. Such local properties enable characterization of asymptotic convergence rates in goodness-of-fit tests when probability measures $P_1$ and $P_0$ are close \cite{Gyorfi}. Although prior research has examined local behaviors of $f$-divergences \cite{Sason}\cite{Bor99}\cite{Vajda0}\cite{Yury1}\cite{Yury4}, quantitative inequalities between pairs of $f$-divergences remain insufficiently explored. 

This work addresses this gap by classifying common $f$-divergences according to their first-order differentiability at $1$. We consider the bound of the $f$-divergences in one class in terms of any one in the other class when the involved pair of probability measures $P_1$ and $P_0$ are close. Our approach is Taylor's Theorem, a methodology established in the literature \cite{Barnett}\cite{Nielsen0}\cite{George1}\cite{George2}. Our analysis employs a generalized quasi-$\varepsilon_{(M,m)}$-neighborhood, which subsumes existing quasi-$\varepsilon$-neighborhood \cite{Huang} while offering three key innovations. Firstly, it extends the existing quasi-$\varepsilon$-neighborhood by providing more flexible parameters $(M,m)$ and offers quantification description on not only common pairs of local distributions with discrete support set, e.g. $\frac{dP_1}{dP_0}$ is close to $1$, all discrete distribution pairs with the same support set can be expressed in this framework with proper chosen parameters. Secondly, by allowing the ratio $\frac{d P_1}{d P_0}$ to be large in some region on the cost that the region should be small in $P_0$ measure, it extends the limit of local behavior characterization in \cite{Vajda0}\cite{Sason1} and  includes several well known continuous distribution pair examples in the literature. For example,  Gaussian local family in \cite{Yury1}\cite{Sason1}, the distribution pair $(1-\lambda)Q + \lambda P$ and $Q$ in \cite{Yury1}\cite{Sason1}. Hence, it provides an unified framework to characterize these pairs of close distributions $P_1$ and $P_0$, which is more manageable and more general than the regular single-parameter families in \cite{Bor99}. Thirdly, it can be applied for more general pairs of $P_1$ and $P_0$ when there is measure concentration in $P_0$,  which includes the truncated exponential families \cite{Nielsen}. Note that, we only provide examples of Gaussian distributions and their truncated versions. Nevertheless, we believe it can be easily extend to other exponential families. 

Within the generalized local setting, we establish bounds between $f$-divergences through the integral form of Taylor's theorem. This approach necessitates third-order differentiability for the second class of $f$-divergences —a stricter condition than the second-order differentiability required in prior works \cite{George1}\cite{George2}. However, this requirement is satisfied by most twice-differentiable $f$-divergences in practice, making our results widely applicable.  Especially, the reverse Pinsker's inequality is re-discovered in this local setting, which includes \cite{Sason}\cite{George1}\cite{George2} in discrete probability space as special case. Note that, it is not our aim to derive bounds of $f$-divergences which outperform the existing results in general settings. We focus on obtaining analytic bounds of $f$-divergences that are easy to compute in a more general local setting and filling the gap in the literature where there are only asymptotic characterizations the ratio of $f$-divergences with $\frac{dP_1}{dP_1}$ approaching $1$. Nevertheless, we provide comparisons between our bounds and the existing sharp bounds in the local setting.

The contributions of our work are summarized as follows: (1) The generalized quasi-$\varepsilon$ neighborhood provides a framework to characterize the closeness between two distributions without structural constraints. (2) We provide inequalities between different $f$-divergences, which generalize the inequalities in terms of $\chi^2$ divergence to characterize mutual asymptotic equivalence in \cite{Vajda0}.  (3) Our results highlight the reverse Pinsker's inequality between total variation distance and other $f$-divergences in a general neighborhood setting, which has not reported in the literature. (4) We provide applications of the inequalities to extend existing asymptotic behavior of $f$-divergence test of goodness of fit. 

\section{Preliminaries}
\subsection{Convex Functions}
Let $(u,v) \subseteq \mathbb{R}$ be a finite or infinite interval, a function $f: (u,v) \mapsto \mathbb{R}$ is convex, then the right derivative and left derivative
\begin{equation}
f'_{+}(s) = \underset{t \downarrow s}{\lim} \frac{f(t) -f(s)}{t-s}, \,  \  \,
f'_{-}(s) = \underset{t \uparrow s}{\lim} \frac{f(t) -f(s)}{t-s}
\end{equation}
always exist and are finite on the whole domain $(u,v)$\cite{Hewitt} \cite{Rockafellar}. Moreover, $f'_{+}(s)$ is always right continuous and monotone nondecreasing. $f'_{-}(s)$ is always left continuous and monotone nondecreasing.
If $f$ is not differentiable at $a$, for $a \leq b$, the Taylor expansion to the convex function $f$ can be written as (Theorem 1 in \cite{Liese})
\begin{equation}
f(b) = f(a) + f'_+(a)(b-a) + R^+_f(a,b)
\end{equation}
where $0 = R_f(a, a) \leq R_f(a,b)$.
While for $a \geq b$, the Taylor expansion to the convex function $f$ can be written as
\begin{equation}
f(b) = f(a) + f'_-(a)(b-a) + R^-_f(a,b)
\end{equation}
for $a,b \in (u,v)$ where $0 = R^-_f(a, a) \leq R^+_f(a,b)$, and
\begin{equation}
R^+_f(a,b) = \int \bm{1}_{(a,b]}(s) (b-s)d f'_-(s)
\end{equation}
\begin{equation}
R^-_f(a,b) = \int \bm{1}_{(b,a]}(s) (s-b)d f'_+(s)
\end{equation}
If a convex function $g(x)$ is twice differentiable at $a$, the Taylor formula can be written as
\begin{equation}
g(b) = g(a) + g'(a)(b-a) + \frac{1}{2}g''(a) (b-a)^2 + R''_g(a,b).
\end{equation}
If $f'''(x)$ exists for every $x \in (a, b)$, then there exists some $\theta \in (a, b)$,
\begin{equation}
R''_g(a,b) = \frac{1}{6}f'''(\theta)(b-a)^3.
\end{equation}

\subsection{$f$-Divergences}
\begin{Definition}
Let $(P_1,P_0)$ be a pair of probability measures defined on a common measure space $(\Omega, \mathcal{F})$ and suppose that $P_1 << P_0$. Given $f: (0,\infty) \rightarrow  \mathbb{R}$ be a convex function such that $f(1) = 0$ . The $f$-divergence from $P_1$ to $P_0$ is given by
\begin{equation}
D_f(P_1\|P_0) = \int f(\frac{dP_1}{dP_0})dP_0
\end{equation}
\end{Definition}
For these $f$, let $f^{\star}: (0,\infty) \rightarrow  \mathbb{R}$ be given by \cite{Liese}
\begin{equation}
f^{\star}(t) = tf(\frac{1}{t}), \,  \, t > 0.
\end{equation}
It is well known that $f^{\star}$ is also convex, $f^{\star}(1) = 0$ and $D_f(P\|Q) = D_{f^{\star}}(Q\|P)$ if $P \ll \gg Q$. Moreover, there is continuous extensions of $f$ and $f^{\star}$ as follows.
\begin{equation}
f(0) = \underset{t\downarrow 0}{\lim} f(t) \in (\infty, \infty],
\end{equation}
\begin{equation}
f^{\star}(0) = \underset{t\downarrow 0}{\lim} f^{\star}(t) = \underset{u \rightarrow \infty}{\lim} \frac{f(u)}{u}.
\end{equation}
Let $f_c(t) = f(t) + c(t -1)$, the following properties are obvious:
(1), $D_{f}(P_1\|P_0) = D_{f_c}(P_1\|P_0)$;
(2), $f(0) + f^{\star}(0) = f_c(0) + f_c^{\star}(0)$.

We list some common $f$-divergences as follows.
\begin{enumerate}
\item [1)] Total variation distance (TV) with $f(t) = \frac{1}{2}|t -1|$:
\begin{equation}
TV(P_1,P_0) = \frac{1}{2}|P_1-P_0| = \underset{A \in \mathcal{F}}{\sup} |P_1(A) - P_0(A)|.
\end{equation}
\item [2)] Kullback-Leibler divergence (KL divergence) with $f(t) = t\log t$:
\begin{equation}
D(P_1\|P_0) = \mathbb{E}_{P_1} \left[\log \frac{dP_1}{dP_0}\right].
\end{equation}

 The relationship between KL divergence and TV can be characterized by  Pinsker's inequality
\begin{equation}
TV(P_1, P_0) \leq \sqrt{ \frac{1}{2}D(P_1\| P_0)}.
\end{equation}
\item [3)]$\chi^2$-Divergence with $f(t) = t^2 - 1$:
\begin{equation}
\chi^2(P_1\|P_0) = \int \left(\frac{dP_1}{dP_0} -1 \right)^2 dP_0.
\end{equation}
\item [4)] Relative Entropy ($P_1 \ll \gg P_0$) with $f(t) = - \log t$:
\begin{equation}
D(P_0\|P_1) = \mathbb{E}_{P_0} \left[\log \frac{dP_0}{dP_1}\right].
\end{equation}
\item [5)] Jefferys' Divergence ($P_1 \ll \gg P_0$) with $f(t) = (t-1) \log t$:
\begin{equation}
D_J(P_1\|P_0) =  D(P_1\| P_0) + \bar{D}(P_0\|P_1).
\end{equation}
\item [6)] Hellinger Distance of order $\alpha \in (0,1) \cup (1, \infty)$ with $f_{\alpha} = \frac{t^{\alpha} -1}{\alpha  - 1}$:
\begin{equation}
H_{\alpha}(P_1,P_0) = \frac{1}{1-\alpha} \left(1 - \int (dP_1)^{\alpha}(dP_0)^{1-\alpha}\right).
\end{equation}
Note that $\chi^2$-Divergence is the Hellinger distance of order $2$, and $\frac{1}{2}H_{\frac{1}{2}}$ is usually referred as Square Hellinger distance $h^2_{\varepsilon}$.
\item [7)] Jensen-Shannon Divergence with $f(t) = t\log t - (1 + t) \log \frac{1+t}{2}$:
\begin{equation}
JS(P_1\|P_0) = D(P_1 \| \frac{P_1+ P_0}{2}) + D(P_0 \| \frac{P_1+ P_0}{2}).
\end{equation}
\end{enumerate}

\subsection{Generalized Quasi-$\varepsilon_{(M,m)}$-Neighborhood}
 In this section, we introduce a notion of generalized quasi-$\varepsilon_{(M,m)}$-neighborhood to characterize the closeness between two distributions.

For a pair of probability measures $P_1$ and $P_0$ on a common measurable space $(\Omega,\mathcal{F})$ and a small number $\varepsilon > 0$, for $\bm{x} \in \Omega$, let $ h_{\varepsilon}(\bm{x}) = \frac{1}{\varepsilon } \left[\frac{d P_1}{dP_0} - 1 \right]$.
\begin{Definition}\label{neighborhooddefintion}
For a given $\varepsilon > 0$, the generalized quasi-$\varepsilon_{(M,m)}$-neighborhood of a reference distribution $P_0$ on $\Omega$ is a set of distributions $P_1$ together with a subset $\Pi_{\varepsilon} \subseteq \Omega$ such that the following conditions are satisfied:
\begin{enumerate}
\item [(1)]
On the subset $\Pi_{\varepsilon} $, the function
 $ h_{\varepsilon}(\bm{x}) = \frac{1}{\varepsilon } \left[\frac{d P_1}{dP_0}(\bm{x}) - 1 \right]$ should satisfy
\begin{equation}\label{genneighborhood}
P_0 - a.s. \, \  \  \,-m \leq h_{\varepsilon}(\bm{x}) \leq M
\end{equation}
where $m > 0, M > 0$.
\item [(2)]
There exists some constant $\tilde{c} < 1$ such that
\begin{equation}\label{assumption01}
\int_{\bar{\Pi}_{\varepsilon}} \frac{(dP_1-dP_0)^2}{dP_0} < \tilde{c} \chi^2(P_1\|P_0).
\end{equation}
where $\bar{\Pi}_{\varepsilon}$ is the complement of $\Pi_{\varepsilon}$.
\item [(3)]
For any $f$-divergence $g$ equipped with third order differentiable function on $\mathcal{R}^+$, define
 \begin{equation}\label{Definition33}
 \bar{\Delta}_{\varepsilon} \triangleq \int_{\bar{\Pi}_{\varepsilon}}  h^2_{\varepsilon}(\bm{x}) \int_0^1 g'''\left[1+ \phi \cdot\varepsilon  h_{\varepsilon}(\bm{x}) \right]\cdot (1 - \phi)d \phi d P_0
 \end{equation}
 and 
\begin{equation}\label{Deltadef}
 \Delta = \int h^2_{\varepsilon}(\bm{x})dP_0,
 \end{equation}
then, there exist some constants $\hat{c}, \hat{c}'$ such that 
\begin{equation}\label{assumption02}
 \hat{c}' \Delta \leq \bar{\Delta}_{\varepsilon} \leq \hat{c} \Delta.
\end{equation}
\end{enumerate}
\end{Definition}
\begin{Remark}\label{explanationonm}
Note that $\varepsilon m\leq 1$ is needed on the subset $\Pi_{\varepsilon} $ with given $\varepsilon$ for $\frac{dP_1}{dP_0} \geq 0$. Different from the local condition in \cite{Vajda0} where $|\frac{d P_1}{d P_0} -1| =  o(1) $ is considered, here  $M\varepsilon = \frac{d P_1}{d P_0} -1$ may be large than $1$, which makes our definition include more general circumstances. We use generalized quasi-$\varepsilon_{(M,m)}$-neighborhood to emphasize that the distributions $P_1$ and $P_0$ are close, which distinguishes it from the bounds in a general setting (For example, Theorem 1, Theorem 4, Theorem 5 and Theorem 23 in \cite{Verdu2}, and the results in \cite{Vajda1},\cite{Binette}, etc). It also extends the results of asymptotic expressions  (For example, Theorem 9 in \cite{Verdu2}, Lemma 4 in \cite{Sason1}) in terms of bounds. Most importantly, the generalized quasi-$\varepsilon_{(M,m)}$-neighborhood makes the closeness under control for the purpose of obtaining analytic and computable bounds for $f$-divergences. 
\end{Remark}
\begin{Remark}
In Definition \ref{neighborhooddefintion}, there are two conditions (\ref{assumption01}) and (\ref{assumption02}). Both conditions involve the integration of $\bm{x}$ on the set $\bar{\Pi}_{\varepsilon}$. The first one (\ref{assumption01}) involves only the distributions $P_1$ and $P_0$, while the second condition (\ref{assumption02}) involves $f$-divergences $g$ which have third order derivative. 
In (\ref{assumption01}), it is required that the partial integration of $h^2_{\varepsilon}(\bm{x})$ in $\bar{\Pi}_{\varepsilon}$ is less than $\tilde{c}\cdot \Delta$.  The second condition involves both the integration of $h^2_{\varepsilon}(\bm{x})$ and the third order derivative $g'''(t)$.  We can rewrite $\bar{\Delta}_{\varepsilon}$ as
 \begin{equation}\notag
 \begin{split}
 \bar{\Delta}_{\varepsilon} = &\int_{\bar{\Pi}_{\varepsilon}}  h^2_{\varepsilon}(\bm{x}) \int_0^1 g'''\left[1+ \phi \cdot\varepsilon  h_{\varepsilon}(\bm{x}) \right]\cdot (1 - \phi)d \phi d P_0 \\
 = & \int_{\bar{\Pi}_{\varepsilon}}  h^2_{\varepsilon}(\bm{x}) g'''\left[1+ \phi_1 \cdot\varepsilon  h_{\varepsilon}(\bm{x}) \right]\cdot (1 - \phi_1)  d P_0 
 \end{split}
 \end{equation} 
 where $\phi_1 \in (0,1)$.
 As $1 + \varepsilon h_{\varepsilon}(\bm{x}) = \frac{d P_1}{d P_0}$, the term $1+ \phi_1 \cdot\varepsilon  h_{\varepsilon}(\bm{x})(1 -\phi_1)$ is between $1$ and $\frac{d P_1}{d P_0}(\bm{x})$. The term $g'''(1+ \phi_1 \cdot\varepsilon  h_{\varepsilon}(\bm{x}))$ may be large, but it still be bounded for most $g$. Nevertheless, the subset $\bar{\Pi}_{\varepsilon}$ is small so that $\bar{\Delta}_{\varepsilon}$ is bounded between $\hat{c}' \Delta$ and $\hat{c} \Delta$. 
\end{Remark}
\begin{Remark}
 From Definition 2, we have
 \begin{equation}
 \begin{split}
 & \chi^2(P_1\|P_0) \\
 = & \int (\frac{dP_1}{dP_0} -1)^2 dP_0 \\
 =  & \int_{\Pi_{\varepsilon}} (\frac{dP_1}{dP_0} -1)^2 dP_0 + \int_{\bar{\Pi}_{\varepsilon}} (\frac{dP_1}{dP_0} -1)^2  dP_0 \\
 < & \int_{\Pi_{\varepsilon}} \varepsilon^2 h_{\varepsilon}^2(\bm{x}) dP_0 + \tilde{c} \chi^2(P_1\|P_0),
 \end{split}
 \end{equation}
 which leads to 
 \begin{equation}
 (1-\tilde{c}) \chi^2(P_1\|P_0) \leq \varepsilon^2 M^2.
 \end{equation}
 Hence, in the generalized quasi-$\varepsilon_{(M,m)}$-neighborhood of a reference distribution $P_0$, 
 we have 
  \begin{equation}
\chi^2(P_1\|P_0) < \frac{\varepsilon^2 M^2}{1-\tilde{c} }.
 \end{equation}
 If $P_0\{\Pi_{\varepsilon}\}= 0$, $P_1$ and $P_0$ will satisfy $-m\varepsilon < \frac{dP_1}{dP_0} - 1 \leq M \varepsilon $. If we further let $\gamma = \max\{m, M\} = 1$, 
our definition will lead to the following inequality. 
\begin{equation}
\chi^2(P_1\| P_0) = \int \left(\frac{dP_1}{dP_0} -1 \right)^2 dP_0 = \int \varepsilon^2 h^2_{\varepsilon} d P_0 \leq \varepsilon^2.
\end{equation}
 In this special case, our definition corresponds to a subset of the quasi-$\varepsilon$-neighborhood defined in \cite{Huang} where it is defined in a discrete probability space as follows.
\end{Remark}
\begin{Definition}\label{basisdistribution}
For a given $\varepsilon > 0$, the quasi-$\varepsilon$-neighborhood of a reference distribution $P_0(z)$ on a discrete probability space $\mathcal{Z}$ is a set of distributions in a $\chi^2$-divergence ball of $\varepsilon^2$ about $P_0(\bm{x})$, i.e.,$
\mathcal{N}_{\varepsilon}(P_0) \triangleq \{ P_1: \chi^2(P_1\| P_0) \leq \varepsilon^2 \},$
where for distributions $P$ and $Q$ on $\mathcal{Z}$ which satisfies $supp(P) \subseteq supp(Q)$,
\begin{equation}\label{chisquare}
\chi^2(P\|Q)  \triangleq \underset{z \in \mathcal{Z}}{\sum} \frac{(Q(z) - P(z))^2}{Q(z)}.
\end{equation}
\end{Definition}

 \section{Main Results}
 In the following, we consider the inequalities of $f$-divergences between two types of $f$-functions in the situation where the distribution $P_1$ in the generalized quasi-$\varepsilon_{(M,m)}$-neighborhood of $P_0$.

  The first type: $\{ f:$  $f$ is convex on $(0, \infty)$ and $f(1) = 0$. In addition, $f$ has unequal right derivative and left derivative at $t = 1\}$.

  The second type: $\{f:$ $f$ is convex on $(0, \infty)$ and $f(1) = 0$. In addition, $f$ has third order derivative around $t = 1\}$.

  For convenience,  the first type and the second type of $f$-divergences are denoted as $D_1$ and $D_2$, respectively.

\begin{Lemma}
For any third order differential convex function $f$, we assume $f'(1) = f(1) = 0$, we have
\begin{equation}
f(1+u) = \frac{1}{2}f''(1)u^2  + \frac{1}{2}u^2 \int_{0}^{1} f'''(1 + \phi u)(1 - \phi)d \phi \end{equation}
\end{Lemma}
\begin{proof}
From Taylor's Theorem with integral form, we have
\begin{equation}
\begin{split}
f(1+u) = & f(1) + f'(1)u  + \frac{1}{2}f''(1)u^2 \\
+& \frac{1}{2} \int_{1}^{1+u}f'''(t)(1+u-t)^2dt\\
= & \frac{1}{2}f''(1)u^2 + \frac{1}{2} \int_{1}^{1+u}f'''(t)(1+u-t)^2dt.
\end{split}
\end{equation}
Now let $t = 1 + \phi u$ for variable substitution, and we will have the conclusion.

\end{proof}
\begin{Theorem}\label{Theorem 1}
For any $f$-divergence $g \in D_2$ with $g''(1) > 0$ and any distribution $P_1(\bm{x})$ in the generalized quasi-$\varepsilon_{(M,m)}$-neighborhood of $P_0(\bm{x})$ with parameters $\tilde{c}, \hat{c}$ and $\Delta$ in Definition \ref{neighborhooddefintion}.
Define
\begin{equation}\label{Gammasup}
 \Gamma_{sup} = \underset{ \underset{\theta \in (1- m\varepsilon, 1+ M\varepsilon)}{ \eta \in (-m, M)}}{\sup} g'''(\theta)\cdot\eta,
  \end{equation}
  \begin{equation}\label{Gammainf}
  \Gamma_{inf} = \underset{ \underset{\theta \in (1-m\varepsilon, 1+ M\varepsilon)}{ \eta \in (-m, M)}}{\inf} g'''(\theta)\cdot\eta.
  \end{equation}
  Define $\gamma = \max \{M,m\}$, and with given $g$, let
\begin{equation}\label{Gdefintion}
G_g(x_1,x_2)= \left\{
            \begin{array}{ll}
              (1+x_1)g''(1), & \forall t > 0, \,  \,g'''(t)< 0  \\
              (1+x_1) g''(1)+ x_2, & \text{otherwise}.
            \end{array}
          \right.
\end{equation}
For any $c$ such that
\begin{equation}\label{condition1}
c \geq  \frac{\varepsilon \Gamma_{sup}}{3 g''(1)},
\end{equation}
we have the following ``reverse" Pinsker's inequality
\begin{equation}\label{tveq1}
\sqrt{\frac{2D_g(P_1\|P_0)}{G_g(c,\hat{c})}} \leq \varepsilon \leq \frac{2\gamma TV(P_1,P_0)}{(1 - \tilde{c})\Delta}.
\end{equation}
\end{Theorem}
\begin{proof}
Now, denote $\Pi_{\varepsilon} = \{\bm{x}: \, \  \,-m \leq h_{\varepsilon}(\bm{x}) \leq M,  \,  \,  P_0 - a.s.\}$.  Let $u = \frac{dP_1}{dP_0} - 1 = \varepsilon h_{\varepsilon}(\bm{x})$ and $g(\frac{dP_1}{dP_0}) = g(1 + u) = g(1 + \varepsilon h_{\varepsilon}(\bm{x}))$, for any $D_2$ we have
\begin{equation}\label{continuoussituation}
\begin{split}
 &D_g(P_1\|P_0)\\
 = & \int g(\frac{dP_1}{dP_0})dP_0\\
 = &\int\left[g(1) + g'(1) (\frac{dP_1}{dP_0} -1) + \frac{1}{2}g''(1) (\frac{dP_1}{dP_0} -1)^2 \right.\\
 &\left.+  R''_g(\frac{dP_1}{dP_0},1)\right]dP_0\\
 = &  \frac{1}{2} \int_{\Pi_{\varepsilon}} g''(1) (\frac{dP_1}{dP_0} -1)^2 dP_0
 \\ + &\frac{1}{6}\int_{\Pi_{\varepsilon}} g'''(\theta) (\frac{dP_1}{dP_0} -1)^3 dP_0  
  + \frac{1}{2}\int_{\bar{\Pi}_{\varepsilon}} g''(1) (\frac{dP_1}{dP_0} -1)^2 dP_0 \\ + &\frac{1}{2} \int_{\bar{\Pi}_{\varepsilon}} \varepsilon^2 h^2_{\varepsilon}(x) \int_0^1 g'''\left[1+\phi (\frac{dP_1}{dP_0} - 1)\right]\cdot (1 - \phi)d \phi d P_0
   \end{split}
 \end{equation}
 \begin{equation}\label{continuous1}
 \begin{split}
   = &\frac{\varepsilon^2}{2} g''(1) \int_{\Pi_{\varepsilon}}  h^2_{\varepsilon}(\bm{x}) dP_0\\
  +  &\frac{\varepsilon^3}{6}\int_{\Pi_{\varepsilon}} g'''(\theta)  h_{\varepsilon}^3(\bm{x}) dP_0 + \frac{\varepsilon^2}{2} g''(1) \int_{\bar{\Pi}_{\varepsilon}}  h^2_{\varepsilon}(\bm{x}) dP_0 \\
    +&  \frac{\varepsilon^2}{2} \int_{\bar{\Pi}_{\varepsilon}}  h^2_{\varepsilon}(\bm{x}) \int_0^1 g'''\left[1+ \phi \cdot\varepsilon  h_{\varepsilon}(\bm{x}) \right]\cdot (1 - \phi)d \phi d P_0.
 \end{split}
\end{equation}
Define
 \begin{equation}\label{Deltaepdef}
 \Delta_{\varepsilon} = \int_{\Pi_{\varepsilon}} h^2_{\varepsilon}(\bm{x})dP_0,
 \end{equation}
  \begin{equation}\label{Deltadef2}
 \Delta'_{\varepsilon} = \int_{\bar{\Pi}_{\varepsilon}} h^2_{\varepsilon}(\bm{x})dP_0,
 \end{equation}

  We have
  \begin{equation}\label{Deltadef}
  \Delta_{\varepsilon} + \Delta'_{\varepsilon} = \Delta = \int h^2_{\varepsilon}(\bm{x})dP_0.
  \end{equation}
 From the definition of $\Delta'_{\varepsilon}$ and the condition (2) in the Definition \ref{neighborhooddefintion},
\begin{equation}
0 \leq \Delta'_{\varepsilon} \leq \frac{\tilde{c}}{ \varepsilon^2} \chi^2(P_1,P_0) = \tilde{c}\Delta,
\end{equation}
\begin{equation}\label{assumption1}
(1 - \tilde{c}) \Delta \leq \Delta_{\varepsilon} \leq \Delta.
\end{equation}
 The first term and the third term of (\ref{continuous1}) are $\frac{\varepsilon^2}{2} g''(1)\Delta_{\varepsilon}$ and $\frac{\varepsilon^2}{2} g''(1)\Delta'_{\varepsilon}$, respectively.
  For the integration of the second term of (\ref{continuous1}), we have
   \begin{equation}
\int_{\Pi_{\varepsilon}} g'''(\theta) h^3(\bm{x})dP_0
\geq \Delta_{\varepsilon}  \cdot \Gamma_{inf}.
 \end{equation}
 \begin{equation}
\int_{\Pi_{\varepsilon}}  g'''(\theta) h^3(\bm{x})dP_0
 \leq \Delta_{\varepsilon} \cdot\Gamma_{sup}
 \end{equation}
 For the fourth term, as $1 + \varepsilon h_{\varepsilon}(\bm{x}) = \frac{dP_1}{dP_0}(\bm{x})$, $\varepsilon h_{\varepsilon}(\bm{x}) > -1$, we have
 \begin{equation}\label{fthirdorder}
  1 + \theta \cdot\varepsilon  h(\bm{x} > 0, \, \  \  \  \,\forall \theta \in (0,1)
 \end{equation}
Denote
 \begin{equation}\label{Definition3}
 \bar{\Delta}_{\varepsilon} =\int_{\bar{\Pi}_{\varepsilon}}  h^2_{\varepsilon}(\bm{x}) \int_0^1 g'''\left[1+ \theta \cdot\varepsilon  h_{\varepsilon}(\bm{x}) \right]\cdot (1 - \theta)d \theta d P_0.
 \end{equation}
As $\chi^2(P_1,P_0) < \infty$, with given $\varepsilon$, from the condition (3) of Definition \ref{neighborhooddefintion}, the quantity  $P_0(\bar{\Pi}_{\varepsilon})$ is small enough such that
\begin{equation}\label{assumption2}
 \hat{c}' \Delta \leq \bar{\Delta}_{\varepsilon} \leq \hat{c} \Delta,
\end{equation}
From (\ref{fthirdorder}), we further let
\begin{equation}\label{evalutionofbarc}
\left\{
          \begin{array}{ll}
             \hat{c}' =  0, & \forall t > 0, \, \, g'''(t) > 0; \\
                \hat{c} =  0, & \forall t > 0, \,  \,g'''(t)< 0; \\
                \hat{c} = \hat{c}'=  0, & \forall t , \,  \,g'''(t) = 0;
          \end{array}
        \right.
\end{equation}

From (\ref{continuous1}), (\ref{assumption1}) and (\ref{assumption2}), the quantity $D_g(P_1\|P_0)$ has the following bound:
 \begin{equation}\label{upperbound1}
 D_g(P_1\|P_0)\leq \frac{\varepsilon^2}{2}g''(1)\cdot \Delta + \frac{\varepsilon^3}{6}\Delta_{\varepsilon} \cdot\Gamma_{sup} + \frac{\varepsilon^2}{2} \hat{c} \Delta \,
 \end{equation}

For any $c$ such that
\begin{equation}\label{condition1}
c \geq  \frac{\varepsilon \Gamma_{sup}}{3 g''(1)},
\end{equation}
the inequalities (\ref{upperbound1}) is rewritten as
\begin{equation}\label{Dgupper1}
\begin{split}
 D_g(P_1\|P_0)
 \leq & \frac{\varepsilon^2}{2}g''(1)\cdot \Delta +  \frac{c\varepsilon^2}{2} g''(1)\Delta + \frac{\varepsilon^2}{2} \hat{c} \Delta\\
 \leq & \left\{
          \begin{aligned}
            \frac{(1+c)g''(1)}{2}\Delta \varepsilon^2, &  \\
               \forall &t > 0, \,  \,g'''(t)< 0 \\
            \left[(1+c) g''(1)+ \hat{c}\right]& \frac{\Delta \varepsilon^2}{2} ,  \text{otherwise}.
          \end{aligned}
        \right.
 \end{split}
\end{equation}

For $TV(P_1,P_0)$, we have
 \begin{equation}\label{TV_geq0}
 \begin{split}
 & V_T(P_1, P_0) \\
 = & \frac{1}{2}\varepsilon \left[\int_{\{h(\bm{x}) > 0\}\cap \Pi_{\varepsilon}}   h(\bm{x}) dP_0 - \int_{\{h(\bm{x}) < 0\}\cap \Pi_{\varepsilon}} h(\bm{x}) dP_0 \right.\\ 
 + & \left. \int_{\bar{\Pi}_{\varepsilon}}|h(\bm{x})| dP_0 \right]  \\
\geq  &\frac{1}{2}\varepsilon \left[\int_{\{h(\bm{x}) > 0\}\cap \Pi_{\varepsilon}}   h(\bm{x}) dP_0 - \int_{\{h(\bm{x}) < 0\}\cap \Pi_{\varepsilon}} h(\bm{x}) dP_0\right] \\
 \geq  & \frac{1}{2\gamma}\varepsilon \left[\int_{\{h(\bm{x}) > 0\}\cap \Pi_{\varepsilon}}   h^2(\bm{x}) dP_0 +  \int_{\{h(\bm{x}) < 0\}\cap \Pi_{\varepsilon}} h^2(\bm{x}) dP_0 \right]\\
 = & \frac{1}{2\gamma} \varepsilon \int_{\Pi_{\varepsilon}} h^2(\bm{x})dP_0\\
 = &  \frac{\varepsilon}{2\gamma} \Delta_{\varepsilon}\\
\geq &\frac{\varepsilon}{2\gamma}(1-\tilde{c}) \Delta.
 \end{split}
 \end{equation}
Combine the inequalities (\ref{Dgupper1}), (\ref{TV_geq0}) and (\ref{Gdefintion}), we have
\begin{equation}\label{tveq1}
\sqrt{\frac{2D_g(P_1\|P_0)}{G_g(c,\hat{c})}} \leq \varepsilon \leq \frac{2\gamma TV(P_1,P_0)}{(1 - \tilde{c})\Delta}.
\end{equation}
\end{proof}
Consider any two $f$-divergences equipped with functions $g_1$ and $g_2$ in $D_2$. For any distribution $P_1(\bm{x})$ in the generalized quasi-$\varepsilon_{(m,M)}$-neighborhood of $P_0(\bm{x})$ with corresponding parameters, we have the following conclusion. 
\begin{Theorem}\label{Theorem 2}
For any distribution $P_1(\bm{x})$ in the generalized quasi-$\varepsilon_{(m,M)}$-neighborhood of $P_0(\bm{x})$ with $\Delta$, $\tilde{c}$, any two $f$-divergences in $D_2$ with corresponding parameters $\hat{c}_1, \hat{c}'_1$ and $\hat{c}_2, \hat{c}'_2$ in Definition \ref {neighborhooddefintion}. With additional parameters $\Gamma_{sup}^{(1)}$, $\Gamma_{sup}^{(2)}$, $\Gamma^{(1)}_{inf}$ and $\Gamma^{(2)}_{inf}$ defined in  (\ref{Gammasup}) and (\ref{Gammainf}) for $g_1$ and $g_2$, respectively. For any $c_1, \bar{c}_1, c_2, \bar{c}_2$ such that
\begin{equation}\label{condition20}
3\bar{c}_1g_1''(1)\leq \varepsilon  \Gamma^{(1)}_{inf},
\varepsilon \Gamma^{(1)}_{sup} \leq 3 c_1 g_1''(1),
\end{equation}
\begin{equation}\label{condition30}
3\bar{c}_2g_2''(1)\leq \varepsilon  \Gamma^{(2)}_{inf},
\varepsilon \Gamma^{(2)}_{sup}\leq 3 c_2 g_2''(1),
\end{equation}
With given $g$, define \begin{equation}\label{Gdefintion1}
G'_g(x_1,x_2,x_3)= \left\{
            \begin{aligned}
(1+x_1 - x_1x_2)g''(1),\,  \  \  \  \  \  \,&\\
 \forall t > 0, \,  \,&g'''(t) > 0; \\
              (1+x_1- x_1x_2) g''(1)+ x_3, &\, \  \  \, \text{otherwise}.
            \end{aligned}
          \right.
\end{equation}
we have
\begin{equation}\label{ineq0}
\frac{G'_{g_1}(\bar{c}_1,\tilde{c},\hat{c}'_1)}{G_{g_2}(c_2,\hat{c}_2)} \leq \frac{D_{g_1}(P_1\| P_0)}{D_{g_2}(P_1\|P_0)}
\leq \frac{G_{g_1}(c_1,\hat{c}_1)}{G'_{g_2}(\bar{c}_2,\tilde{c},\hat{c}'_2)},
\end{equation}
where the function $G_{g_i},G_{g_i}'$ are defined in (\ref{Gdefintion}) and (\ref{Gdefintion1}) with functions $g_1$ and $g_2$.
\end{Theorem}
\begin{proof}
 With $\Delta$ defined in (\ref{Deltadef}), $\Gamma_{sup}^{(1)}$ and $\Gamma_{sup}^{(2)}$ as defined in (\ref{Gammasup}) for $g_1$ and $g_2$ and $\Gamma^{(1)}_{inf}$ and $\Gamma^{(2)}_{inf}$ defined in (\ref{Gammainf}) for $g_1$ and $g_2$. 
  From (\ref{continuous1}), (\ref{assumption1}) and (\ref{assumption2}), the quantity $D_{g_1}(P_1\|P_0)$ has the following bound:
 \begin{equation}\label{lowerbound1}
\begin{split}
 &D_{g_1}(P_1\|P_0)\\
 =  &\frac{\varepsilon^2}{2} g_1''(1) \int_{\Pi_{\varepsilon}}  h^2_{\varepsilon}(\bm{x}) dP_0\\
  +  &\frac{\varepsilon^3}{6}\int_{\Pi_{\varepsilon}} g_1'''(\theta)  h^3(\bm{x}) dP_0 + \frac{\varepsilon^2}{2} g_1''(1) \int_{\bar{\Pi}_{\varepsilon}}  h^2_{\varepsilon}(\bm{x}) dP_0 \\
    +&  \frac{\varepsilon^2}{2} \int_{\bar{\Pi}_{\varepsilon}}  h^2_{\varepsilon}(\bm{x}) \int_0^1 g_1'''\left[1+ \theta \cdot\varepsilon  h_{\varepsilon}(\bm{x}) \right]\cdot (1 - \theta)d \theta d P_0 \\
&\geq \frac{\varepsilon^2}{2}g_1''(1)\cdot \Delta + \frac{\varepsilon^3}{6}\Delta_{\varepsilon} \cdot\Gamma^{(1)}_{inf} + \frac{\varepsilon^2}{2} \hat{c}_1' \Delta.
\end{split}
 \end{equation}
 From the first inequality in (\ref{condition20}), 
the inequality (\ref{lowerbound1}) is rewritten as
\begin{equation}\label{lowerfinal0}
\begin{split}
 D_{g_1}(P_1\|P_0)
 \geq &\, \, \frac{\varepsilon^2}{2}g_1''(1)\cdot \Delta +  \frac{\bar{c}_1\varepsilon^2}{2} g_1''(1)\Delta_{\varepsilon} + \frac{\varepsilon^2}{2} \hat{c}_1' \Delta\\
\geq & \, \, \frac{\varepsilon^2}{2}g_1''(1)\cdot \Delta +  \frac{\bar{c}_1\varepsilon^2}{2} g_1''(1)(1- \tilde{c})\Delta + \frac{\varepsilon^2}{2} \hat{c}_1' \Delta\\
\geq & \left\{
          \begin{aligned}
(1+\bar{c}_1 - \bar{c}_1\tilde{c})g''(1)\frac{\Delta \varepsilon^2}{2},& \\  \forall t > 0, &\,  \,g'''(t) > 0; \\
            \left[(1+\bar{c}_1- \bar{c}_1\tilde{c}) g_1''(1)+ \hat{c}'\right] &\frac{\Delta \varepsilon^2}{2},  \text{otherwise}.
          \end{aligned}
        \right.\\
=  & \, \  \, G'_{g_1}(\bar{c}_1,\tilde{c},\hat{c}_1')\cdot \frac{\Delta \varepsilon^2}{2}.
\end{split}
\end{equation}
Together with (\ref{Dgupper1}) for $g_1$ with parameters $c_1$ and $\hat{c}_1$, we have
\begin{equation}\label{D_g1geq1}
 \begin{split}
G'_{g_1}(\bar{c}_1,\tilde{c},\hat{c}_1')\cdot \frac{\Delta \varepsilon^2}{2} \leq D_{g_1}(P_1\|P_0) \leq G_{g_1}(c_1,\hat{c}_1)\cdot \frac{\Delta \varepsilon^2}{2}
 \end{split}
 \end{equation}
The same derivation applies for $g_2$ will lead to 
\begin{equation}\label{D_g1geq2}
 \begin{split}
G'_{g_2}(\bar{c}_2,\tilde{c},\hat{c}'_2)\cdot \frac{\Delta \varepsilon^2}{2} \leq D_{g_2}(P_1\|P_0) \leq G_{g_2}(c_2,\hat{c}_2)\cdot \frac{\Delta \varepsilon^2}{2}
 \end{split}
 \end{equation}
Thus, the inequality (\ref{ineq0}) is obvious from (\ref{D_g1geq1}) and (\ref{D_g1geq2}). 
\end{proof}
 \begin{Remark}
 All these parameters $c,\bar{c}, \bar{c}'$ and $\tilde{c}$ are presumed to be small, which are correct for some typical applications. In addition, when $P_0(\bar{\Pi}_{\varepsilon}) \rightarrow 0$, we have $c,\bar{c}, \bar{c}'$ and $\tilde{c}$ tend to $0$, and hence $G_g(c,\hat{c}) \rightarrow 
 g''(1)$ and $G'_{g}(\bar{c},\tilde{c},\hat{c}') \rightarrow g''(1)$. Hence, in the asymptotic sense, we have
 \begin{equation}
\begin{split}
 \frac{g_1''(1)}{g_2''(1)} \leftarrow & \max \, \  \, \frac{G'_{g_1}(\bar{c}_1,\tilde{c},\hat{c}'_1)}{G_{g_2}(c_2,\hat{c}_2)} \\
   \leq &\frac{D_{g_1}(P_1\| P_0)}{D_{g_2}(P_1\|P_0)}
\leq \min\, \ \, \frac{G_{g_1}(c_1,\hat{c}_1)}{G'_{g_2}(\bar{c}_2,\tilde{c},\hat{c}'_2)} \rightarrow  \frac{g_1''(1)}{g_2''(1)} .
 \end{split}
\end{equation}
 \end{Remark}
 
In Theorem 1 and Theorem 2, although it is required that the distribution $P_1$ is in a generalized quasi-$\varepsilon_{\gamma}$ (or $\varepsilon_{(m,M)}$) neighborhood of $P_0$, with given $\varepsilon $ such that $\varepsilon m \leq 1$, the value of $\gamma$ (or $M$ ) can be large, which implies that the two distributions may not be very close. The sufficient conditions for these inequalities in both theorems depend on the conditions of the inequalities (\ref{condition1}), (\ref{condition2}) and (\ref{condition3}), which further depend on the value domain of the ratio $\frac{dP_1}{dP_0}$ and the properties of related $f$-functions. With given $\varepsilon$ and $\gamma$, the evaluation of $\Gamma_{sup}$ and $\Gamma_{inf}$ may be complicated when dealing with some specific divergences. Nevertheless, our derivation provides a general framework to bound a $f$-divergence in terms of another, which can be utilize to characterize the quantity relationship between various $f$-divergences. The bounds are available when the underlying probability measures has the phenomenon of measure concentration. Especially, when $P_0$ is a probability measure with concentration in subset $\Pi_{\varepsilon} $, and $P_1$ is the empirical distribution of the samples from $P_0$, the conditions (1) (2) and (3) in Definition \ref{neighborhooddefintion} are easy to meet, which will illustrated by examples in Section \ref{applicationspart}. 

The next two corollaries focus on the special situation wherew $P_0(\Pi_{\varepsilon}) = 1$. A typical scenario is that $P_1$ and $P_0$ are discrete distributions with the same support set. In the case, $P_1$ is expressed as $\bm{p} = (p_1,\cdots,p_n)$ and $P_0$ is expressed as $\bm{q} = (q_1,\cdots, q_n)$.
\begin{Corollary}\label{localcondition}
For any $f$-divergence $g \in D_2$ with $g''(1) > 0$, any distribution $P_1(\bm{x})$ in generalized quasi-$\varepsilon_{M,m}$-neighborhood of $P_0(\bm{x})$ such that $P_0(\Pi_{\varepsilon}) = 1$. Let 
$\Gamma_{sup}$,$\Gamma_{inf}$ and $\Delta$ be defined in ( \ref{Gammasup}) (\ref{Gammainf}) and (\ref{Deltadef}), respectively. 
For any $c$ such that
\begin{equation}\label{condition1}
c \geq  \frac{\varepsilon \Gamma_{sup}}{3 g''(1)},
\end{equation}
and $\gamma = \max\{m, M\} $,
we have
\begin{equation}\label{neighborhoodin}
TV(P_1, P_0) \geq \frac{1}{2\gamma}\Delta\varepsilon, \, \, D_g(P_1\| P_0) \leq \frac{1+ c}{2}\varepsilon^2 g''(1)\Delta .
\end{equation}
Moreover, we can get the following ``reverse" Pinsker's inequality:
\begin{equation}\label{inequality3}
 \sqrt{\frac{\Delta D_g(P_1\|P_0)}{2(1+c)g''(1)\gamma^2}} \leq TV(P_1, P_0).
\end{equation}
\end{Corollary}
\begin{proof}

For the second type which has continuous third order derivative around $t = 1$, we have
 \begin{equation}\label{secondtype}
 \begin{split}
 &D_g(P_1\|P_0)\\
 = & \int g(\frac{dP_1}{dP_0})dP_0\\
 = &\int\left[g(1) + g'(1) (\frac{dP_1}{dP_0} -1) + \frac{1}{2}g''(1) (\frac{dP_1}{dP_0} -1)^2 \right.\\
 &\left.+  R''_g(\frac{dP_1}{dP_0},1)\right]dP_0\\
\end{split}
\end{equation}
\begin{equation}\nonumber
\begin{split}
 =  &\int\left[ \frac{1}{2}g''(1) (\frac{dP_1}{dP_0} -1)^2 + \frac{1}{6}g'''(\theta) (\frac{dP_1}{dP_0} -1)^3\right]dP_0\\
 = & \frac{\varepsilon^2}{2} g''(1)  \int h^2_{\varepsilon}(\bm{x})dP_0 + \frac{\varepsilon^3}{6}\int g'''(\theta) h^3(\bm{x})dP_0 \\
 \end{split}
 \end{equation}
 where $ \theta \in (1- m \varepsilon, 1 +  M\varepsilon)$ and it depends on the sample $\bm{x}$.
  We have the following inequalities for the term involving the third order derivative from (\ref{Gammasup}), (\ref{Gammainf}) and (\ref{Deltadef}),
 \begin{equation}
\int g'''(\theta) h^3(\bm{x})dP_0
\geq \Delta  \cdot \Gamma_{inf},
 \end{equation}
 \begin{equation}
\int g'''(\theta) h^3(\bm{x})dP_0
 \leq \Delta \cdot\Gamma_{sup}.
 \end{equation}
 Then, we arrive at the following bound for $D_g$:
\begin{equation}\label{D_gobunds1}
\begin{split}
\frac{1}{2}\varepsilon^2 g''(1) \Delta  +  \frac{1}{6}\varepsilon^3  \Delta \Gamma_{inf} & \leq D_g(P_1\|P_0) \\
 & \leq  \frac{1}{2}\varepsilon^2 g''(1) \Delta + \frac{1}{6}\varepsilon^3 \Delta \Gamma_{sup}.
\end{split}
\end{equation}
Since $g''(1) > 0$, for any $c$ such that
$\varepsilon \Gamma_{sup}  \leq 3 c g''(1)$,
we further have
\begin{equation}\label{D_gleq}
 \begin{split}
D_g(P_1\|P_0) \leq \frac{1+ c}{2}\varepsilon^2 g''(1) \Delta.
 \end{split}
 \end{equation}
 The following derivation is almost the same as (\ref{TV_geq0}). 
 \begin{equation}\label{TV_geq}
 \begin{split}
 & TV(P_1, P_0) \\
 = & \frac{1}{2}\varepsilon \left[\int_{h_{\varepsilon}(\bm{x}) > 0}   h_{\varepsilon}(\bm{x}) dP_0 - \int_{h_{\varepsilon}(\bm{x}) < 0} h_{\varepsilon}(\bm{x}) dP_0\right] \\
 \geq  & \frac{1}{2\gamma}\varepsilon \left[\int_{h_{\varepsilon}(\bm{x}) > 0}   h^2_{\varepsilon}(\bm{x}) dP_0 +  \int_{h_{\varepsilon}(\bm{x}) < 0} h^2_{\varepsilon}(\bm{x}) dP_0 \right]\\
 = & \frac{1}{2\gamma} \varepsilon \int h^2_{\varepsilon}(\bm{x})dP_0\\
 = &  \frac{1}{2\gamma} \Delta \varepsilon.
 \end{split}
 \end{equation}
Combine the inequalities (\ref{D_gleq}) and (\ref{TV_geq}), we finally have
\begin{equation}
\sqrt{\frac{2D_g(P_1\|P_0)}{ g''(1)\Delta (1+c)}} \leq \varepsilon \leq \frac{2\gamma TV(P_1,P_0)}{\Delta}.
\end{equation}
\end{proof}
\begin{Remark}\label{remarklabel1}
 Note that $\Delta$ and $\gamma$ only depend on the distributions $P_1$ and $P_0$, while $c$ depends on both the distribution pair and the divergence $g$. From the definition of $h_{\varepsilon}(\bm{x})$, we have
 \begin{equation}\label{rewritten1}
 \Delta = \int h^2_{\varepsilon}(\bm{x})dP_0 = \frac{1}{\varepsilon^2} \int \left(\frac{d P_1}{d P_0}-1\right)^2 d P_0 = \frac{\chi^2(P_1\|P_0)}{\varepsilon^2},
 \end{equation}
 and the quantity $\frac{\Delta}{\gamma^2}$ can be rewritten as
 \begin{equation}\label{rewritten2}
 \frac{\Delta}{\gamma^2} = \int \left(\frac{h_{\varepsilon}(\bm{x})}{\gamma}\right)^2 d P_0 = \frac{\chi^2(P_1\|P_0)}{(\varepsilon\gamma)^2}.
 \end{equation}
 From (\ref{rewritten1}), $\Delta$ measures the size of the generalized quasi-$\varepsilon$-neighborhood respect to the standard quasi- $\varepsilon$-neighborhood. From (\ref{genneighborhood}) and $\gamma = \max \{m,M\}$, the middle term of (\ref{rewritten2}) is less than $1$, hence $\frac{\Delta}{\gamma^2}$ is a normalized coefficient. Furthermore, the inequalities in (\ref{neighborhoodin}) can be rewritten as
 \begin{equation}\label{directchi1}
\chi^2(P_1\|P_0)\leq  2\gamma \varepsilon \cdot TV(P_1, P_0) 
 \end{equation}
 and
 \begin{equation}\label{directchi3}
 D_g(P_1\|P_0) \leq \frac{1+c}{2} g''(1)\chi^2(P_1\|P_0).
 \end{equation}
 The inequality (\ref{directchi3}) is equivalent as the second part of the inequality of Theorem 31 in \cite{George2}. 
 The inequality (\ref{directchi1}) is tighter than the inequality (33) in \cite{George2} in the discrete case where $P_1$ has the form of $\bm{p} = (p_1,\cdots,p_n)$ and $P_0$ has the form of $\bm{q} = (q_1,\cdots, q_n)$. From the definition of $h_{\varepsilon}(x)$, we have $1 + \gamma \varepsilon = \max{\frac{p_i}{q_i}}$. The coefficients of $TV(P_1,P_0)$ at the right-hand side of the inequality (33) in \cite{George2} and the inequality (\ref{directchi1}) are $\frac{\|\bm{p} - \bm{q}\|}{\min q_i}$ and $\max \frac{p_i}{q_i} -1 = \max \frac{p_i - q_i}{q_i}$, respectively. From the fact that 
 \begin{equation}
 \frac{\max |p_i - q_i|}{\min q_i} \geq \max \frac{p_i - q_i}{q_i}, 
 \end{equation}
our bound in (\ref{directchi1}) is tighter than that in \cite{George2}. 
 \end{Remark}

 For $\Gamma_{sup}$ and $\Gamma_{inf}$, we have the following detailed calculation procedures, which will be useful in next section.
  \begin{enumerate}
  \item [(1)] If $g'''(\theta)  \equiv 0$, we have
  \begin{equation}\label{Gammaeq0}
  \Gamma_{sup} =    \Gamma_{inf} = 0.
  \end{equation}
  \item [(2)] If $\forall \theta \in (1-m\varepsilon, 1+ M\varepsilon)$, $g'''(\theta)> 0$, we have
  \begin{equation}\label{Gammasup1}
  \Gamma_{sup} = M \cdot\underset{\theta \in (1-m\varepsilon, 1+ M\varepsilon)}{\sup}g'''(\theta)
   \end{equation}and
   \begin{equation}\label{Gammainf1}
   \Gamma_{inf} = - m \cdot \underset{\theta \in (1-m\varepsilon, 1+ M\varepsilon)}{\sup}g'''(\theta).
   \end{equation}
  \item [(3)] If $\forall \theta \in (1-m\varepsilon, 1+ M\varepsilon)$, $g'''(\theta)< 0$, we have
  \begin{equation}\label{Gammasup2}
  \Gamma_{sup} = -m \cdot \underset{\theta \in (1-m\varepsilon, 1+ M\varepsilon)}{\inf}g'''(\theta)
   \end{equation}and
   \begin{equation}\label{Gammainf2}
   \Gamma_{inf} = M \cdot \underset{\theta \in (1-m\varepsilon, 1+ M\varepsilon)}{\inf}g'''(\theta)
   \end{equation}
  \end{enumerate}
 If $\gamma \rightarrow 0$, then $m \rightarrow 0, M\rightarrow 0$. Furthermore, from (\ref{Gammasup1}) to (\ref{Gammainf2}), it further results in
 \begin{equation}\label{asymptoticbehaviorgamma}
 \Gamma_{sup} \downarrow 0,\, \  \  \  \,  \Gamma_{inf} \uparrow 0.
 \end{equation}

\begin{Corollary} \label{differentablef}
For any two $f$-divergences equipped with functions $g_1$ and $g_2$ in $D_g$ and any distribution $P_1(\bm{x})$ in the generalized quasi-$\varepsilon_{(m,M)}$-neighborhood of $P_0(\bm{x})$ such that $P_0(\Pi_{\varepsilon}) = 1$. With $\Delta$ defined in (\ref{Deltadef}), $\Gamma_{sup}^{(1)}$ and $\Gamma_{sup}^{(2)}$ as defined in (\ref{Gammasup}) for $g_1$ and $g_2$ and $\Gamma^{(1)}_{inf}$ and $\Gamma^{(2)}_{inf}$ defined in (\ref{Gammainf}) for $g_1$ and $g_2$. For any $c_1, \bar{c}_1, c_2, \bar{c}_2$ such that
\begin{equation}\label{condition2}
3\bar{c}_1g_1''(1)\leq \varepsilon  \Gamma^{(1)}_{inf},
\varepsilon \Gamma^{(1)}_{sup} \leq 3 c_1 g_1''(1),
\end{equation}
\begin{equation}\label{condition3}
3\bar{c}_2g_2''(1)\leq \varepsilon  \Gamma^{(2)}_{inf},
\varepsilon \Gamma^{(2)}_{sup}\leq 3 c_2 g_2''(1),
\end{equation}
we have
\begin{equation}\label{ineq0}
\frac{(1+\bar{c}_1)g_1''(1)}{(1+c_2)g_2''(1)} \leq \frac{D_{g_1}(P_1\| P_0)}{D_{g_2}(P_1\|P_0)}
\leq \frac{(1+c_1)g_1''(1)}{(1+\bar{c}_2)g_2''(1)}.
\end{equation}
\end{Corollary}
\begin{proof}
Consider the inequalities (\ref{D_gobunds1}) for $g_1$, from the conditions in (\ref{condition2}), we have
\begin{equation}\label{D_ggeq}
 \begin{split}
D_{g_1}(P_1\|P_0) \geq \frac{1+ \bar{c}_1}{2}\varepsilon^2 g_1''(1) \Delta,
 \end{split}
 \end{equation}
and
\begin{equation}\label{g_1upp}
D_{g_1}(P_1\|P_0)
  \leq  \frac{1+c_1}{2}\varepsilon^2 g_1''(1) \Delta.
\end{equation}
Similarly, for $g_2$ we have
\begin{equation}\label{D_ggeq2}
 \begin{split}
D_{g_2}(P_1\|P_0) \geq \frac{1+ \bar{c}_2}{2}\varepsilon^2 g_2''(1) \Delta,
 \end{split}
 \end{equation}
and
\begin{equation}\label{g_1upp2}
D_{g_2}(P_1\|P_0)
  \leq  \frac{1+c_2}{2}\varepsilon^2 g_2''(1) \Delta.
\end{equation}
Thus, from (\ref{g_1upp}) and (\ref{D_ggeq2}),
 \begin{equation}
 \frac{D_{g_1}(P_1\|P_0)}{(1 + c_1)g''_1(1)} \leq \frac{1}{2}\varepsilon^2 \Delta \leq \frac{D_{g_2}(P_1\|P_0)}{(1+\bar{c}_2)g''_2(1)}
 \end{equation}
 which is equivalent to the right side of (\ref{ineq0}).
After swapping the role of $g_1$ and $g_2$, we have the left side of (\ref{ineq0}).
\end{proof}
Note that the inequalities in (\ref{ineq0}) are nontrivial only if $1 + \bar{c}_1 > 0$ and $1 + \bar{c}_2 > 0$, which means with given $m,M$, we should choose $\varepsilon$ with caution.
\begin{Remark}
Recently, the following inequality is provided in Theorem 31 of \cite{George2}:
\begin{equation}\label{Geogeineq}
\frac{\kappa_f^{\uparrow}(\bm{p},\bm{q})}{2} \chi^2(\bm{p}\|\bm{q}) \leq D_f(P_1\|P_0) \leq \frac{\kappa_f^{\downarrow}(\bm{p},\bm{q})}{2} \chi^2(\bm{p}\|\bm{q})
\end{equation}
where $\kappa_f^{\uparrow}(\bm{p},\bm{q})$ and $\kappa_f^{\downarrow}(\bm{p},\bm{q})$ are expressed as (24) and (25) in \cite{George2}. The above inequality (\ref{Geogeineq}) are equivalent to the inequalities from (\ref{D_ggeq}) to (\ref{g_1upp2}) because the parameters $\varepsilon$ and $m,M$ can be easily determined in the discrete case. In other words, we can always find some generalized quasi-$\varepsilon_{M,m}$-neighborhood to incorporate the cases of discrete probabilities. Hence, the established bounds in Theorem \ref{Theorem 1} and Theorem \ref{Theorem 2} are more general than Theorem 31 in \cite{George2}. 
\end{Remark}

 \begin{Remark}
 The following inequality has been reported in \cite{Basu} and \cite{Verdu2}:
 \begin{equation}\label{fdivergencelinear}
 D_f(P\|Q) \leq (f(0) + f^{\star}(0))TV(P,Q).
 \end{equation}
 where $f$ is the function equipped with $f$-divergence. Consider the $f$-divergence in left side of (\ref{fdivergencelinear}) in $D_g$, and we write it as $D_g(P_1\|P_0)$ and $TV(P_1,P_0)$ for convenience. From (\ref{fdivergencelinear}), we get a lower bound of $TV(P_1,P_0)$ as a linear function of $D_f$,
 \begin{equation}\label{TVlowerbound1}
 \frac{1}{g(0) + g^{\star}(0)} D_g(P_1\|P_0) \leq TV(P_1,P_0).
 \end{equation}
 Compare the two lower bounds of $TV(P_1\|P_0)$ from (\ref{TVlowerbound1}) and (\ref{inequality3}), we have
 \begin{equation}
   \frac{1}{g(0)+ g^{\star}(0)} D_g(P_1\|P_0) \leq \sqrt{\frac{\Delta D_g(P_1\|P_0)}{2(1+c)g''(1)\gamma^2}},
  \end{equation}
  \begin{equation}
 \iff  \frac{D_g(P_1\|P_0)}{\Delta} \leq \frac{(g(0)+g^{\star}(0))^2}{2(1+c)g''(1)\gamma^2},
 \end{equation}
 \begin{equation}\label{tighter1}
 \iff \varepsilon ^2 \cdot \frac{D_g(P_1\|P_0)}{\chi^2(P_1\|P_0)} \leq \frac{(g(0)+g^{\star}(0))^2}{2(1+c)g''(1)\gamma^2}.
 \end{equation}
 From (\ref{directchi3}), the left side of (\ref{tighter1}) is upper bounded by $\frac{1+c}{2} g''(1)\varepsilon^2 $. Hence, the inequality (\ref{tighter1}) holds if
 \begin{equation}
 \frac{1+c}{2} g''(1) \varepsilon^2 \leq \frac{[g(0)+g^{\star}(0)]^2}{2(1+c)g''(1)\gamma^2}.
 \end{equation}
It is equivalent to
 \begin{equation}
 |\varepsilon (1+c)g''(1) \gamma | \leq |g(0) + g^{\star}(0)|,
 \end{equation}
 which obviously holds if $\varepsilon$ is small. Thus, when $P_1$ and $P_0$ is sufficiently close so that $P_1$ is in $\varepsilon_{\gamma}$-neighborhood and the condition (\ref{tighter1}) is satisfied, the lower bound of TV in terms of $f$-divergence in $D_g$ in (\ref{inequality3}) is tighter than (\ref{TVlowerbound1}).

 In \cite{Binette}, it is proved that
 \begin{equation}\label{fdivergencelinear2}
 \underset{(P,Q) \in \mathcal{A}(\delta, \hat{m}, \hat{M})}{\sup} D_f(P\|Q) = \delta \left(\frac{f(\hat{m})}{1-\hat{m}} + \frac{f(\hat{M})}{\hat{M}-1}\right)
 \end{equation}
 In the above formula (\ref{fdivergencelinear2}), we have $\hat{m} = 1 - \varepsilon m$ and $\hat{M} = 1 +\varepsilon M$. For the same reason as above, the inequality (\ref{inequality3}) provides a refined lower bound of TV in terms of square root of $f$-divergence with $f$ in type $D_g$  in a generalized $\varepsilon_{\gamma}$-neighborhood.
 \end{Remark}
\begin{Remark}
From Theorem 5 and (156) in \cite{Verdu2}, related upper and lower bounds are
\begin{equation}\label{generalizedlinearbound}
\begin{split}
\underset{\beta \in (\beta_2, 1)\cup (1, \beta_1^{-1})}{\inf} \kappa (\beta)& D_{g_2}(P\|Q)
\leq  D_{g_1}(P\|Q) \\
\underset{\beta \in (\beta_2, 1)\cup (1, \beta_1^{-1})}{\sup} \kappa(\beta) &D_{g_2}(P\|Q) \geq D_{g_1}(P\|Q).
 \end{split}
\end{equation}
where $g_1,g_2$ can be any $f$-functions (not confined to the type $D_g$).
When the function $\kappa$ is monotonically increasing, the inequalities in (\ref{generalizedlinearbound}) are converted into
\begin{equation}\label{generalizedlinearbound2}
\kappa(\beta_2) D_g(P \|Q) \leq D_f(P\|Q) \leq \kappa(\beta_1^{-1}) D_g(P\|Q).
\end{equation}
Compared with (\ref{generalizedlinearbound}) and (\ref{generalizedlinearbound2}), our bounds focus on the situation when both $\beta_1, \beta_2$ \footnote{We have $\beta^{-1}_1 = 1 + \varepsilon M$, $\beta_2 = 1-\varepsilon m$ in our work}are relatively close to $1$ and we proceed to relate the supremum and infimum of $\kappa(\beta)$ to the second and third order derivatives of $g_1,g_2$. Especially, the coefficients $c_i, \bar{c}_i, i= 1,2$ depend on both $\beta_1, \beta_2$ and the second and third derivatives of $g_1, g_2$. In addition, our bounds in Theorem \ref{differentablef} don't rely on the monotonicity of $\kappa$, and thus can provide bounds of more pairs of $f$-divergences than the inequality (\ref{generalizedlinearbound2}).
\end{Remark}
\begin{Remark}\label{Remark4}
From Theorem \ref{differentablef}, tighter bounds can be found as
\begin{equation}\label{ineq00}
\begin{split}
\underset{\bar{c}_1, c_2}{\max}& \frac{(1+\bar{c}_1)g_1''(1)}{(1+c_2)g_2''(1)} D_{g_2}(P_1\|P_0)\\
 \leq & D_{g_1}(P_1\| P_0) \\
\leq & \underset{c_1, \bar{c}_2}{\min} \frac{(1+c_1)g_1''(1)}{(1+\bar{c}_2)g_2''(1)} D_{g_2}(P_1\|P_0).
\end{split}
\end{equation}
When $\gamma \rightarrow 0$, from (\ref{asymptoticbehaviorgamma}), (\ref{condition2}) and (\ref{condition3}), we have $\bar{c}_1g_1''(1)\leq \frac{\varepsilon \Gamma^1_{inf}}{3} \rightarrow 0$, $c_1 g_1''(1) \geq \frac{\varepsilon \Gamma^1_{sup}}{3}\rightarrow 0$, $\bar{c}_2g_2''(1)\leq \frac{\varepsilon \Gamma^2_{inf}}{3}\rightarrow 0$ and $c_2 g_2''(1) \geq \frac{\varepsilon \Gamma^2_{sup}}{3} \rightarrow 0$. From the inequality (\ref{ineq0}), we have
\begin{equation}
\begin{split}
 \frac{g_1''(1)}{g_2''(1)} \leftarrow &\underset{\bar{c}_1,c_2}{\max} \, \  \, \frac{g_1''(1) + \bar{c}_1g_1''(1)}{g_2''(1) +  c_2 g_2''(1)}\\
  \leq &\frac{D_{g_1}(P_1\| P_0)}{D_{g_2}(P_1\| P_0)} \\
   \leq &\underset{c_1,\bar{c}_2}{\min}\, \ \, \frac{g_1''(1) + c_1g_1''(1)}{g_2''(1) +  \bar{c}_2 g_2''(1)} \rightarrow \frac{g_1''(1)}{g_2''(1)}.
 \end{split}
\end{equation}
Hence, the inequalities can be regarded as intermediate results between the limit behavior of Lemma 4 in \cite{Sason1} and the bounds in (\ref{generalizedlinearbound}). When applying Theorem \ref{differentablef} for specific $f$-divergence pairs (1): $D(P_1\|P_0)$ and $D(P_0\|P_1)$; (2): $D(P_1\|P_0)$ and $\chi^2(P_1\|P_0)$, the results indicate the limits indicated by (181) of Corollary 2 and (182) of Corollary 3.
\end{Remark}

\section{Some Applications} \label{applicationspart}
In this section, we provide some applications of our bounds with some typical distributions. These distributions have emerged in previous literature. 
\subsection{Local Family}
Let $P_0 \sim \mathcal{N}(0,1)$ and $P_t \sim \mathcal{N}(t, 1)$ where $t \in \mathbb{R}$ that is close to the origin \cite{Yury1}. In this case, we know that
    \begin{equation}
    TV(P_t, P_0) = \frac{|t|}{(2\pi)^{\frac{1}{2}}}
    \end{equation}
    and if $g$ be a twice continuously differentiable convex function such that $D_g$ is a $f$-divergence and  $\chi^2(P\|Q) < \infty$, then
    \begin{equation}
    D_g(P_t\|P_0) = \frac{1}{2}g''(1)t^2 + o(t^2).
    \end{equation}
    Especially, 
    \begin{equation}
    \chi^2(P_t\|P_0) = e^{t^2}-1.
    \end{equation}
    For the quantity $\frac{dP_t}{dP_0} - 1$, we have
\begin{equation}
    \begin{split}
    \frac{dP_t}{dP_0} - 1 = &\frac{\frac{1}{\sqrt{2\pi}}e^{-\frac{(x-t)^2}{2}}}{\frac{1}{\sqrt{2\pi}}e^{-\frac{x^2}{2}}} - 1 
     =  e^{tx - \frac{t^2}{2}} - 1,
    \end{split}
    \end{equation}
With given $\varepsilon, M, m$, for $t > 0$ close to $0$, we have
\begin{equation}
e^{\frac{1}{2}tx} -1 \leq e^{tx - \frac{t^2}{2}} - 1, \  \  \  \  \, \text{for}\  \  \   x > 0
\end{equation}
and 
\begin{equation}
 e^{tx - \frac{t^2}{2}} - 1 \leq e^{tx} -1, \  \  \  \  \, \text{for}\  \  \   x < 0
\end{equation}
and thus
\begin{equation}
\{x: x > \frac{2 \ln(1 + M \varepsilon)}{t} \} \subseteq  \{x:\frac{dP_t}{dP_0}(x) - 1 > M\varepsilon  \} ,
\end{equation}
\begin{equation}
\{x: x < \frac{\ln(1 - m \varepsilon)}{t} \} \subseteq \{x:\frac{dP_t}{dP_0}(x) - 1 <  - m\varepsilon  \}.
\end{equation}
In this case, the subset $\bar{\Pi}_{\varepsilon}$ should satisfy
\begin{equation}
A_{\varepsilon} \triangleq \{x: x > \frac{2\ln(1 + M \varepsilon)}{t} \, \, \text{or}  \, \, x < \frac{\ln(1 - m \varepsilon)}{t}\} \subseteq \bar{\Pi}_{\varepsilon} 
\end{equation}
Using the inequality
\begin{equation}
\frac{1}{\sqrt{2\pi}} \frac{x}{x^2+ 1}e^{-\frac{x^2}{2}} < \mathbb{P}(Z > x)< \frac{1}{\sqrt{2\pi}} \frac{1}{x}e^{-\frac{x^2}{2}} 
\end{equation}
where $Z$ is a standard Gaussian random variable and $x > 0$. From (\ref{assumption01}), a proper $\tilde{c}$ is a lower bound of $\frac{\int_{A_{\varepsilon}}{(\frac{dP_t}{dP_0}-1)^2} dP_0}{e^{t^2} -1}$ as
\begin{equation}\label{localinequality1}
\begin{split}
 \tilde{c} = & \frac{1}{2\pi}\cdot \frac{1}{e^{t^2}-1} \left[ \frac{2t M^2\varepsilon^2\ln(1+ M\varepsilon)}{t^2+ 4\ln^2(1 + M\varepsilon)} +  \frac{t m^2\varepsilon^2\ln(1 -m\varepsilon)}{t^2+ 4\ln^2(1 -m\varepsilon)}\right]
\end{split}
\end{equation}
For simplicity, let's consider the pair of divergences: TV and $\chi^2$-divergence, the latter as a $g$ function $g(t) = t^2 - 1$. As $g'''(t) =0$, from (\ref{Gammasup}) and (\ref{evalutionofbarc}), we have $c = \hat{c} = 0$. From Theorem 1, we have
\begin{equation}
\sqrt{\chi^2(P_t\|P_0)} \leq \frac{2\gamma TV(P_t,P_0)}{(1-\tilde{c})\Delta}.
\end{equation}
The bound between TV and other divergences can be derived similarly. 
\subsection{$(1-\lambda)Q + \lambda P$}  Consider linear combination of two measures $P$ and $Q$ in the form of $(1-\lambda)Q + \lambda P$ where $P$ can be arbitrary measure such that $\chi^2(P\|Q) < \infty$ \cite{Yury1}\cite{Sason1}. For this combination, a statement is as follows \cite{Yury1}.
    Let $f$ be a twice continuously differentiable convex function such that $D_f$ is a $f$-divergence, then $D_g((1-\lambda)Q + \lambda P ||Q) < \infty$ for all $0 < \lambda < 1$. Moreover, for $\lambda$ close to $0$,
    \begin{equation}\label{temp1}
    D_g((1-\lambda)Q + \lambda P \|Q) = \frac{g''(1)}{2} \chi^2(P\|Q) + o(\lambda^2).
    \end{equation}
From
    \begin{equation}\label{temp1}
    \frac{d [(1-\lambda)Q + \lambda P ]}{d Q} = 1 -\lambda + \lambda \frac{dP}{dQ},
    \end{equation}
    we have $\frac{d [(1-\lambda)Q + \lambda P ]}{d Q}$ can be arbitrarily large with any fixed $\lambda$. From (\ref{temp1}), we get
\begin{equation}
\begin{split}
\{\bm{x}:\frac{d[(1-\lambda)Q + \lambda P ]}{dP_0}(\bm{x}) - 1 > M\varepsilon  \} \\
= \{\bm{x}: \frac{dP}{dQ}(\bm{x}) > 1 + \frac{M\varepsilon}{\lambda} \},
\end{split}
\end{equation}
and
\begin{equation}
\begin{split}
\{\bm{x}:\frac{d[(1-\lambda)Q + \lambda P ]}{dP_0}(\bm{x}) - 1 < - m\varepsilon  \} \\
= \{\bm{x}: \frac{dP}{dQ}(\bm{x}) < 1 - \frac{m\varepsilon}{\lambda} \}.
\end{split}
\end{equation}
If  $\gamma' = \min\{M,m\}$, we have $\bar{\Pi}_{\varepsilon} \subset \{ \bm{x}: |\frac{d P}{dQ}(\bm{x}) -1| > \frac{\gamma' \varepsilon}{\lambda}\}$. By Markov'a inequality, 
\begin{equation}
\chi^2(P\|Q) = \int(\frac{d P}{dQ} -1)^2 dQ  \geq  \frac{\gamma'^2 \varepsilon^2}{\lambda^2} P( |\frac{d P}{dQ}(\bm{x}) -1| > \frac{\gamma' \varepsilon}{\lambda}),
\end{equation}
we arrive at
\begin{equation}
P(\bar{\Pi}_{\varepsilon} )\leq P( |\frac{d P}{dQ}(\bm{x}) -1| > \frac{\gamma' \varepsilon}{\lambda}) \leq \frac{\lambda^2\chi^2(P\|Q)}{\gamma'^2 \varepsilon^2}
\end{equation}
Hence, $P_0(\bar{\Pi}_{\varepsilon})$ is small with sufficiently small $\lambda$ such that the conditions (2) and (3) in Definition 1 are satisfied, so the inequality (\ref{tveq1}) and (\ref{ineq0}) hold for the distribution pair $(Q,(1-\lambda)Q + \lambda P)$. From (\ref{assumption01}), a proper choice of $\tilde{c}$ is 
\begin{equation}
\tilde{c} = \frac{\lambda^2\chi^2(P\|Q) }{\chi^2((1-\lambda)Q + \lambda P \|Q)}.
\end{equation}
Consider the pair of divergences: TV and $\chi^2$-divergence, we have
\begin{equation}
\sqrt{\chi^2((1-\lambda)Q + \lambda P\|Q)} \leq \frac{2\gamma TV((1-\lambda)Q + \lambda P,Q)}{(1-\tilde{c})\Delta}
\end{equation}
where $\gamma = \max \{M,m\}$.
\subsection{Gaussian Distributions and Their Truncated Versions}
\subsubsection{Low Dimension Situation}
Consider Gaussian distribution $P^n_0\sim \mathcal{N}(\bm{0}, P\bm{I}_n)$ on $\mathbb{R}^n$ with small $n$ ($n \leq 20$), whose density function is
\begin{equation}\label{gdensity}
g^{(n)}(\bm{x}) = \frac{1}{(2\pi  P)^{n/2}}e^{-\frac{\|\bm{x}\|^2}{2 P}}.
\end{equation}
The truncated Gaussian distribution $P_1^n$ with density function
\begin{equation}\label{fdensity}
f^{(n)}(\bm{x}) = \left\{
\begin{split}
&\frac{1}{\Theta}g^{(n)}(\bm{x}),    \,  \  \  \  \  \  \  \  \  \  \  \  \, \|\bm{x}\|\hspace{-0.04in}\leq \hspace{-0.04in} \sqrt{yP},\\
&0,    \ \  \  \  \  \  \  \  \ \  \  \  \  \  \  \ \ \  \  \  \  \  \  \  \  \,otherwise.
\end{split}
\right.
\end{equation}
In the above formula, $\Theta$ is the normalized coefficient
 \begin{equation}
 \Theta = E_{P_0^n}[1_{\{\bm{x} \in \mathfrak{B}^n_0(  \sqrt{yP})\}}],
 \end{equation}
 where $\mathfrak{B}^n_0(r)$ is the $n$-dimensional sphere with center $\bm{0}$ and radius $r$.
 We have
 \begin{equation}\label{trunction}
TV(P_0^n, P_1^n) = \frac{1}{2}\underset{\mathcal{R}^n}{\int}|f^{(n)}(\bm{x}) - g^{(n)}(\bm{x}))|d\bm{x} = 1- \Theta.
\end{equation}
From theorem 1 in \cite{Jurg}, when $y$ is sufficiently large, we have the tail probability satisfies
\begin{equation}\label{tailprobabilitygaussian}
P_0^n(\|\bm{x}\| > \sqrt{Py})= 1 - \Theta  \sim \frac{2^{1-\frac{n}{2}}}{\Gamma(\frac{n}{2})} e^{-\frac{y}{2}}y^{\frac{n}{2}-1}.
\end{equation}
It implies that $1- \Theta$ tends to $0$ exponentially with increasing $y$ and fixed $n$.
 \begin{equation}
 \frac{f_n(\bm{x})}{g_n(\bm{x})} - 1 =
 \left\{
\begin{split}
&\frac{1}{\Theta}-1,    \,  \  \  \  \  \  \  \  \  \  \  \  \  \ \,|\bm{x}\|\hspace{-0.04in}\leq \hspace{-0.04in}\sqrt{yP},\\
&-1,    \ \  \  \  \  \  \  \  \ \ \  \  \  \  \  \  \  \,otherwise
\end{split}
\right.
 \end{equation}
 From the neighborhood condition (1), we have
\begin{equation}\label{condition1Gaussianhigh}
\begin{split}
 & -\varepsilon m \leq\frac{1}{\Theta} - 1 \leq M \varepsilon\\
\iff & \frac{1}{1+ M\varepsilon} \leq \Theta \leq \frac{1}{1-\varepsilon m} \\
\iff & -\frac{\varepsilon m}{1-\varepsilon m} \leq 1-\Theta \leq \frac{M\varepsilon}{1 + \varepsilon M}.
\end{split}
\end{equation}
From (\ref{tailprobabilitygaussian}), we have $\mathfrak{B}^n_0(\sqrt{yP}) = \Pi_{\varepsilon}$ with sufficiently large $y$. Thus, we have $P^n_0(\bar{\Pi}_{\varepsilon}) = 1 - \Theta$.  The inequality (\ref{assumption01}) for $\tilde{c}$ is expressed as 
\begin{equation}
\tilde{c} > \frac{P_0^n(\|\bm{x}\| > \sqrt{Py})}{\chi^2(P_1^n \| P_0^n) } = \frac{1-\Theta}{\Theta \cdot \frac{(1-\Theta)^2}{\Theta^2} + (1 -\Theta)} = \Theta.
\end{equation}

\subsubsection{High Dimension - Sphere Hardening Effect}

\begin{figure*}
\centering
\includegraphics[width=6in]{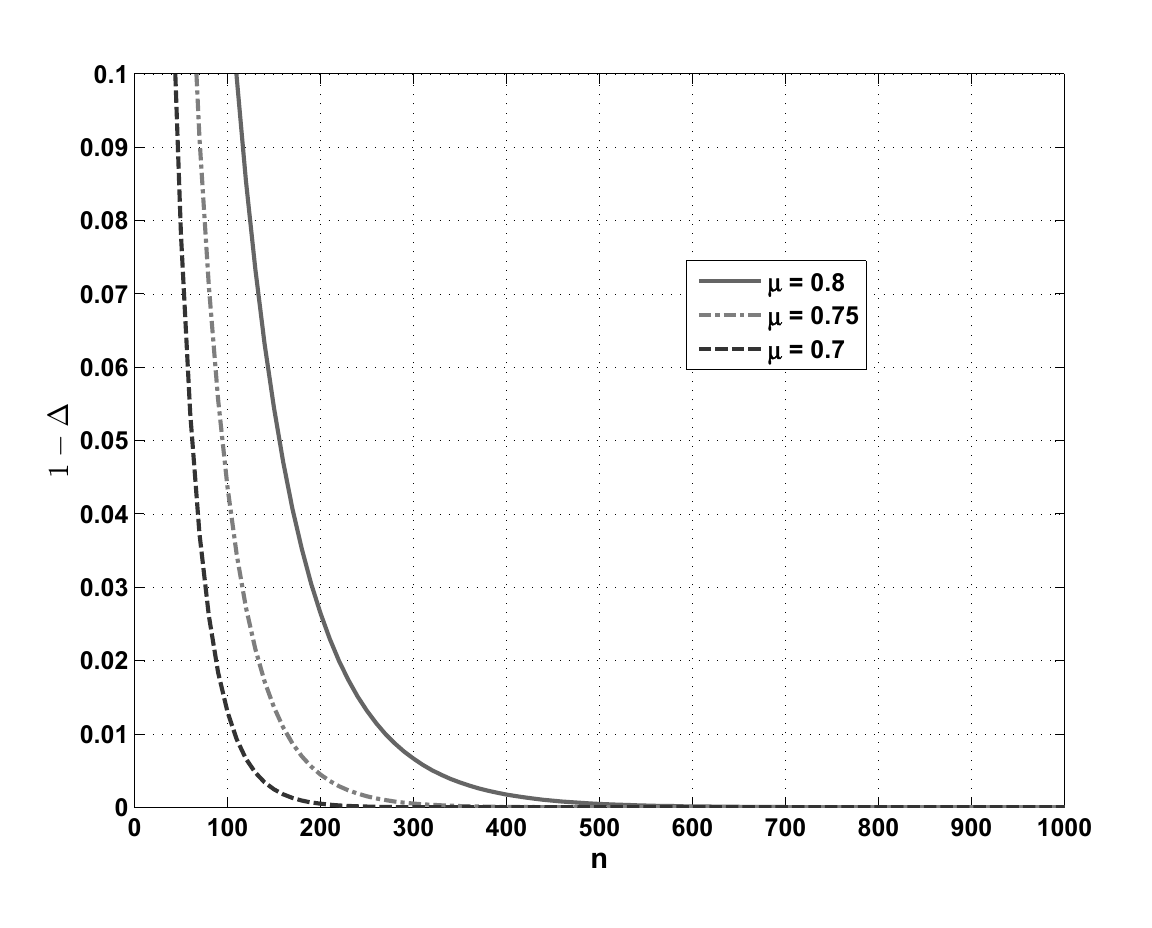}
\caption{$TV(\bar{P}_0^n, \bar{P}_1^n)$ between Gaussian distributions and their truncated versions.}\label{Fig1}
\end{figure*}

Consider Gaussian distribution $\bar{P}^n_0\sim \mathcal{N}(\bm{0}, \mu P\bm{I}_n)$ on $\mathbb{R}^n$ with $0 < \mu < 1$ and large $n$ ($n \geq 100$), whose density function is
\begin{equation}\label{gdensity}
\bar{g}^{(n)}(\bm{x}) = \frac{1}{(2\pi \mu P)^{n/2}}e^{-\frac{\|\bm{x}\|^2}{2\mu P}}.
\end{equation}
 The truncated Gaussian distribution $P_1^n$ with density function \cite{Yu1}
\begin{equation}\label{fdensity}
\bar{f}^{(n)}(\bm{x}) = \left\{
\begin{split}
&\frac{1}{\Delta}g^{(n)}(\bm{x}),    \sqrt{\mu^2 nP} \leq \|\bm{x}\|\hspace{-0.04in}\leq \hspace{-0.04in}  \sqrt{nP},\\
&0,    \ \  \  \  \  \  \  \  \ \  \  \  \  \  \  \  \  \  \  \  \  \  \  \  \  \  \  \,otherwise.
\end{split}
\right.
\end{equation}
In the above formula, $\Delta$ is the normalized coefficient
 \begin{equation}
 \Delta = E_{\bar{P}_0^n}[1_{\{\bm{x} \in \mathfrak{B}^n_0(\sqrt{nP})\backslash \mathfrak{B}^n_0(\sqrt{\mu^2nP})\}}].
 \end{equation}
We have
 \begin{equation}\label{trunction}
TV(\bar{P}_0^n, \bar{P}_1^n) = \frac{1}{2}\underset{\mathcal{R}^n}{\int}|\bar{f}^{(n)}(\bm{x}) - \bar{g}^{(n)}(\bm{x})|d\bm{x} = 1- \Delta.
\end{equation}
After integrate $r$ from $0$ to $\sqrt{nP}$, we have
\begin{equation}
\Delta =  \frac{\gamma(n/2, n/2\mu)- \gamma(n/2, n\mu/2)}{\Gamma(n/2)}
\end{equation}
where $\gamma(a,z)$ is incomplete gamma function defined as follows
 \begin{equation}
 \gamma(a, z)= \int_0^{z}e^{-t}t^{a-1}dt.
 \end{equation}
 As $\Delta$ tends to $1$ exponentially fast for increasing $n$ and fixed $\mu$ close to 1, see Fig.\ref{Fig1}. It is known as sphere hardening effect \cite{Hamkins}.
 we have
 \begin{equation}
 \frac{\bar{f}_n(\bm{x})}{\bar{g}_n(\bm{x})} - 1 =
 \left\{
\begin{split}
&\frac{1}{\Delta}-1,    \sqrt{\mu^2nP} \leq \hspace{-0.04in}\|\bm{x}\|\hspace{-0.04in}\leq \hspace{-0.04in}\sqrt{nP},\\
&-1,    \ \  \  \  \  \  \  \  \ \ \  \  \  \  \  \  \  \,otherwise
\end{split}
\right.
 \end{equation}
From the neighborhood condition (1), we have
\begin{equation}\label{condition1Gaussianhigh}
\begin{split}
 & -\varepsilon m \leq\frac{1}{\Delta} - 1 \leq M \varepsilon\\
\iff & \frac{1}{1+ M\varepsilon} \leq \Delta \leq \frac{1}{1-\varepsilon m} \\
\iff & -\frac{\varepsilon m}{1-\varepsilon m} \leq 1-\Delta \leq \frac{M\varepsilon}{1 + \varepsilon M}
\end{split}
\end{equation}
From the fact that the value $\frac{1}{\Delta}-1$ is close to $0$ with $\mu$ close to $1$, the condition (\ref{condition1Gaussianhigh}) holds and the subset $\Pi_{\varepsilon} = \mathfrak{B}^n_0(\sqrt{nP})\backslash \mathfrak{B}^n_0(\sqrt{\mu^2nP})$.
Thus, we have $P^n_0(\bar{\Pi}_{\varepsilon}) = 1 - \Delta$. The inequality (\ref{assumption01}) for $\tilde{c}$ is expressed as 
\begin{equation}
\tilde{c} > \frac{\bar{P}_0^n(\bm{x} \notin \mathfrak{B}^n_0(\sqrt{nP})\backslash \mathfrak{B}^n_0(\sqrt{\mu^2nP}))}{\chi^2(\bar{P}_1^n \| \bar{P}_0^n) } = \Delta.
\end{equation}

\section{Applications to Specific $f$-divergences}

In this section, we will consider the conditions on specific functions $g_1$ and $g_2$ under which the inequalities in Corollary \ref{localcondition} and Corollary \ref{differentablef} hold, i.e. the inequalities (\ref{condition1}), (\ref{condition2}) and (\ref{condition3}). For both theorems, we list some applications to some particular $f$-divergences which belong to the second type $D_g$. The obtained bounds are compared to the similar inequalities shown in the literature.
\subsection{$f$-Divergence Inequalities in Discrete Probability Space}
\subsubsection{$D(P_1\|P_0)$ and $TV(P_1,P_0)$}
For $D(P_1\|P_0)$ with $f(t) = t\log t$, we have
$f''(t) = \frac {\log e}{t}$ and $f'''(t) = - \frac{1}{t^2} \log e$. Since $f'''(\theta) < 0$, from (\ref{Gammasup2}) we get
\begin{equation}\label{GD1}
\begin{split}
 \Gamma_{sup} = & -m \cdot \underset{\theta \in (1-m\varepsilon, 1+ M\varepsilon)}{\inf}g'''(\theta)
= \frac{m}{(1- m \cdot\varepsilon)^2} \log e.
\end{split}
\end{equation}
From (\ref{condition1}), the qualified $c$ for the inequality (\ref{inequality3}) should satisfy
\begin{equation}\label{reversec}
c \geq \frac{\varepsilon \Gamma_{sup}}{3g''(1)} = \frac{\varepsilon m}{3(1 - \varepsilon m)^2}.
\end{equation}
Thus, we have the following bounds of $TV(P_1,P_0)$ in terms of $D(P_1\|P_0)$ from Theorem \ref{localcondition} and Pinsker's inequality:
\begin{equation}\label{lowerboundKL1}
\begin{split}
\sqrt{\frac{\Delta D(P_1\|P_0)}{2(1+c)\gamma^2 \log e}} & = \sqrt{\frac{\chi^2(P_1\|P_0) \ln 2}{2(1+c) (\gamma \varepsilon)^2}D(P_1\|P_0)}\\
& \leq TV(P_1, P_0)\\
 &\leq  \sqrt{ \frac{1}{2}D(P_1\| P_0)}.
\end{split}
\end{equation}
The tightest bound for (\ref{lowerboundKL1}) by taking equality for the inequality of
 $c$ in (\ref{reversec}).
The ratio between the lower bound and the upper bound in (\ref{lowerboundKL1}) is expressed as
\begin{equation}\label{ratiobetweenbounds}
\frac{\chi^2(P_1\|P_0)}{ (\gamma \varepsilon)^2} \cdot \frac{\ln 2}{(1+c)} = \frac{\Delta}{\gamma^2} \cdot \frac{\ln 2}{(1+c)},
\end{equation}
which is the normalized coefficient multiplied by a constant.
Especially, when $P_1(\bm{x})$ is in the $\varepsilon_{\gamma}$-neighborhood of $P_0(\bm{x})$ with $\varepsilon = 0.1$ and $\gamma  = 1$, the inequality (\ref{condition1}) is satisfied by $c = 1$. Thus, we have
\begin{equation}\label{inequalities1}
\begin{split}
\frac{1}{2}\sqrt{D(P_1\| P_0)\Delta \ln 2} \leq TV&(P_1, P_0)
 \leq  \sqrt{ \frac{1}{2}D(P_1\| P_0)}.
\end{split}
\end{equation}
The inequalities (\ref{lowerboundKL1}) and (\ref{inequalities1}) provide direct quantitative relationships between $\sqrt{D(P_1\| P_0)}$ and $TV(P_1, P_0)$ as reverse Pinsker's inequalities in local settings.
\begin{Remark}
Recently,  the following lower bounds of TV in terms of KL divergence have been found.
\begin{enumerate}
\item
 In Theorem 23 of \cite{Verdu2}, a linear lower bound of TV in terms of KL divergence is given as \footnote{In Theorem 23 and Theorem 25 of \cite{Verdu2}, $|P-Q| = 2 TV(P,Q)$. }
 \begin{equation}\label{gobalsettinglinear}
 D(P\|Q) \leq  \left( \varphi(\beta^{-1}_1) - \varphi(\beta_2)TV(P,Q)\right),
 \end{equation}
 where $\varphi$ is given by
 \begin{align}
 \varphi(t) =
 \begin{cases}
0 & t =0; \\
\frac{t\log t}{t-1} & t\in (0,1)\cup (1, \infty);\\
\log e &t = 1.
 \end{cases}
 \end{align}
 From (\ref{gobalsettinglinear}) and (\ref{lowerboundKL1}),
 \begin{align}
 \frac{1}{\varphi(\beta_1^{-1}) - \varphi(\beta_2)} D(P_1\|P_0) \leq \sqrt{\frac{\Delta D(P_1\|P_0)}{2(1+c)\gamma^2\log e}}\\
 \iff D(P_1\|P_0) \leq \frac{(\varphi(\beta_1^{-1}) - \varphi(\beta_2))\Delta}{2(1+c)\gamma^2\log e}. \label{tightercondition}
 \end{align}
 When $P_1$ and $P_0$ is close so that (\ref{tightercondition}) holds, the bound (\ref{lowerboundKL1}) is a tighter lower bound than (\ref{gobalsettinglinear}), which will be further illustrated in Section \ref{Examplesection}. 
 \item In the case of finite alphabet, it is shown in Theorem 25 of \cite{Verdu2} that
 \begin{equation}\label{finitealphabet}
 D(P\|Q) \leq  \log \left(1 + \frac{2 TV(P,Q)^2}{Q_{min}}\right).
 \end{equation}
 \end{enumerate}
The left hand side of (\ref{inequalities1}) is a refinement of (\ref{gobalsettinglinear}) in the form of square root of $D(P_1\| P_0)$ in a local setting, and is more straightforward than (\ref{finitealphabet}). 
\end{Remark}
From (\ref{inequalities1}), we can see that $\Delta$ is important for the quantity relationship between KL divergence and TV when the two distributions are close.
\subsubsection{$\chi^2(P_1\|P_0)$ and $TV(P_1,P_0)$}
We consider the specific pair from the inequality (\ref{inequality3}). For $\chi^2$-Divergence with $g(t) = t^2 - 1$, we have
$g''(t) = 2$ and $g'''(t) = 0$.  Hence,
$\Gamma_{sup} = 0$.
From (\ref{condition1}), the qualified $c$ for the inequality (\ref{inequality3}) should satisfy
$c \geq 0$. Let $c = 0$, we have
\begin{equation}\label{chi2VT}
\frac{1}{2}\sqrt{\frac{\Delta \chi^2(P_1\|P_0)}{\gamma^2}} \leq TV(P_1,P_0)
\end{equation}
From (\ref{rewritten1}) in Remark \ref{remarklabel1}, the above inequality is rewritten as
\begin{equation}\label{directchi2}
\begin{split}
TV(P_1,P_0) & \geq  \frac{1}{2} \sqrt{\frac{ \chi^2(P_1\|P_0)}{\gamma^2} \cdot \frac{\chi^2(P_1\|P_0)}{\varepsilon^2}}\\
 & = \frac{\chi^2(P_1\|P_0)}{2 \varepsilon \gamma},
\end{split}
\end{equation}
which is the same as (\ref{directchi1}).
Especially, when $P_1(\bm{x})$ is in the generalized $\varepsilon_{\gamma}$-neighborhood of $P_0(\bm{x})$ with $\gamma  = 1$, the inequality (\ref{condition1}) is satisfied by $c = 0$. Thus, we have
\begin{equation}
 TV(P_1, P_0) \geq \frac{1}{2}\sqrt{\chi^2(P_1\| P_0)\Delta } = \frac{\chi^2(P_1\|P_0)}{2 \varepsilon},
\end{equation}
which is the same as the inequality (\ref{directchi2}) with $\gamma = 1$.
\begin{Remark}
In (159) of \cite{Verdu2}, a related lower bound of TV in terms of $\chi^2(P\|Q)$ is
\begin{equation}\label{chi2TV}
\chi^2(P_1\|P_0) \leq 2 \cdot \max\{\beta_1^{-1} - 1, 1 - \beta_2\} \cdot TV(P_1,P_0).
\end{equation}
In the generalized $\varepsilon_{\gamma}$-neighborhood, we have $\beta_1 = \frac{1}{1+ \varepsilon M}$ and $\beta_2 = 1 - \varepsilon m$, hence we have
\begin{equation}
\max \{\beta_1^{-1} -1, 1-\beta_2\} = \max \{\varepsilon M, \varepsilon m\} = \varepsilon \gamma
\end{equation}
and the inequality (\ref{directchi2}) is equivalent to (\ref{chi2TV}) in this local setting.

\end{Remark}

\subsubsection{$H_{\alpha}(P_1,P_0)$ and $TV(P_1,P_0)$}
For Hellinger distance of order $\alpha$, it is easy to get $g''(t) = \alpha t^{\alpha - 2}$ and $g'''(t) = \alpha (\alpha -2)t^{\alpha - 3}$. For $\varepsilon \gamma \leq 1$, the range $(1-\varepsilon m, 1 + \varepsilon M)$ lies on the positive half axis, and whether $g'''(\theta)$ is positive or not only depends on  $\alpha$. Thus, we have from (\ref{Gammasup1}) and (\ref{Gammasup2}):
\begin{flalign}
 \Gamma_{sup} &= \underset{ \underset{\theta \in (1-\varepsilon m,1+ \varepsilon M)}{ \kappa \in (-m, M)}}{\sup} g'''(\theta)\eta \\
 & =  \underset{ \underset{\theta \in (1-\varepsilon m,1+ \varepsilon M)}{ \eta \in (-m, M)}}{\sup} \eta\alpha (\alpha -2)\theta^{\alpha - 3}
 \end{flalign}
 \begin{equation}
 = \left\{
 \begin{aligned}
        &-m \times \underset{\theta \in (1-\varepsilon m,1+ \varepsilon M)}{\inf} \alpha (\alpha -2)\theta^{\alpha - 3}, \\
           &\,  \  \  \  \  \  \  \  \  \  \  \  \  \  \  \  \  \  \  \  \  \  \  \  \  \  \  \  \  \  \  \  \  \  \  \  \  \, 0 < \alpha < 2, \alpha \neq 1; \\
    &0 \,  \  \  \  \  \  \  \  \  \  \  \  \  \  \  \  \  \  \  \  \  \  \  \  \  \  \  \  \  \  \  \  \  \  \  \  \  \  \  \  \  \  \  \  \  \  \  \  \  \,  \alpha = 2; \\
      &M \times \underset{\theta \in (1-\varepsilon m,1+ \varepsilon M)}{\sup} \alpha (\alpha -2)\theta^{\alpha - 3}, \, \  \ \,2 < \alpha < 3;\\
      &M \times \underset{\theta \in (1-\varepsilon m,1+ \varepsilon M)}{\sup} \alpha (\alpha -2)\theta^{\alpha - 3},   \, \ \  \  \  \  \  \  \, \alpha \geq 3 \\
 \end{aligned}
    \right.
    \end{equation}
     \begin{equation}
 = \left\{
 \begin{aligned}
        &-m \alpha (\alpha -2)\times \underset{\theta \in (1-\varepsilon m,1+ \varepsilon M)}{\sup} \frac{1}{\theta^{3-\alpha }}, \\
           &\,  \  \  \  \  \  \  \  \  \  \  \  \  \  \  \  \  \  \  \  \  \  \  \  \  \  \  \  \  \  \  \  \  \  \  \  \  \, 0 < \alpha < 2, \alpha \neq 1; \\
               &0 \,  \  \  \  \  \  \  \  \  \  \  \  \  \  \  \  \  \  \  \  \  \  \  \  \  \  \  \  \  \  \  \  \  \  \  \  \  \  \  \  \  \  \  \  \  \  \  \  \  \,  \alpha = 2; \\
      &M  \alpha (\alpha -2)\times \underset{\theta \in (1-\varepsilon m,1+ \varepsilon M)}{\sup}\frac{1}{\theta^{3-\alpha }}, \, \  \,2 < \alpha < 3;\\
      &M\alpha (\alpha -2) \times \underset{\theta \in (1-\varepsilon m,1+ \varepsilon M)}{\sup} \theta^{\alpha - 3},   \, \ \  \  \  \  \  \  \, \alpha \geq 3 \\
 \end{aligned}
    \right.
    \end{equation}
    \begin{equation}\label{gammasupHellinger}
    \begin{split}
= &  \begin{cases}
        -m \cdot \frac{\alpha(\alpha -2)}{(1 - \varepsilon m)^{3-\alpha}}, & 0 < \alpha < 2, \alpha \neq 1; \\
       0, & \alpha  = 2; \\
     M \cdot \frac{\alpha(\alpha -2)}{(1 - \varepsilon m)^{3-\alpha}}, & 2 < \alpha < 3;\\
     M \cdot \alpha(\alpha -2)(1 + \varepsilon M)^{\alpha- 3}, &  \alpha \geq 3.\\
      \end{cases}
\end{split}
\end{equation}
The qualified region of $c$ for the inequality (\ref{inequality3}) can be calculated from (\ref{condition1}) as follows.
\begin{equation}\label{cVTHalpha}
\begin{split}
c \geq  \frac{\varepsilon \Gamma_{sup}}{3 g''(1)}
= & \frac{\varepsilon \Gamma_{sup}}{3 \alpha}\\
= &  \begin{cases}
     \frac{-\varepsilon m(\alpha -2)}{3(1 - \varepsilon m)^{3-\alpha}}, & 0 < \alpha < 2, \alpha \neq 1; \\
       0, & \alpha  = 2; \\
     \frac{\varepsilon M(\alpha -2)}{3(1 - \varepsilon m)^{3-\alpha}}, & 2 < \alpha < 3;\\
 \frac{\varepsilon M(\alpha -2)(1 + \varepsilon M)^{\alpha- 3}}{3}, &  \alpha \geq 3.\\
      \end{cases}
\end{split}
\end{equation}
We have the following bound of $TV(P_1,P_0)$ in terms of $H_{\alpha}(P_1,P_0)$ from Theorem \ref{localcondition}:
\begin{equation}\label{HalphaVTbound}
\sqrt{\frac{\Delta H_{\alpha}(P_1\|P_0)}{2\alpha(1+c)\gamma^2}} \leq TV(P_1,P_0).
\end{equation}
The tightest bound for (\ref{HalphaVTbound}) by taking equalities for the inequalities of
 $c$ in (\ref{cVTHalpha}).

Consider some special case. For example, when $\alpha = 2$, Hellinger distance of order $2$ is $\chi^2$-Divergence and $c \geq 0$, the inequality (\ref{HalphaVTbound}) becomes (\ref{chi2VT}).
When $P_1(\bm{x})$ is in the generalized quasi-$\varepsilon_{\gamma}$-neighborhood of $P_0(\bm{x})$ with $\varepsilon = 0.1$, $\gamma  = 1$ and $0 <\alpha < 2, \alpha \neq 1$, the inequality (\ref{condition1}) is satisfied by $c = 1$. Thus, we have
\begin{equation}\label{inequalities2}
\frac{1}{2}\sqrt{\frac{\Delta}{\alpha}H_{\alpha}(P_1\| P_0)} \leq TV(P_1, P_0).
\end{equation}

\subsubsection{$D(P_1\|P_0)$ and $D(P_0\|P_1)$}
Let $g_1(t) =  t\log t$ and $g_2(t) = -\log t$, it is easy to get $g''_1(t) = \frac{\log e}{t}$, $g''_2(t) = \frac{\log e}{t^2}$, $g'''_1(t) = -\frac{1}{t^2}\log e$ and $g'''_2(t) = -\frac{2}{t^3}\log e$. For $\varepsilon \gamma \leq 1$, we have $g'''_1(\theta) < 0$ and $g'''_2(\theta) < 0$ for $\theta \in (1-\varepsilon m, 1+ \varepsilon M)$. From (\ref{Gammasup2}) and (\ref{Gammainf2}), we get
\begin{equation}\label{GammaKLsup}
\begin{split}
\Gamma^{(1)}_{sup} = & -m \cdot \underset{\theta \in (1-m\varepsilon, 1+ M\varepsilon)}{\inf}\left[ -\frac{1}{\theta^2}\log e\right]\\
= & m \cdot \underset{\theta \in (1-m\varepsilon, 1+ M\varepsilon)}{\sup} \frac{1}{\theta^2} \log e\\
= & m \cdot \frac{\log e}{\underset{\theta \in (1-m\varepsilon, 1+ M\varepsilon)}{\inf} \theta^2}\\
= &\frac{m \log e}{(1 - \varepsilon m)^2},
\end{split}
\end{equation}
\begin{equation}\label{GammaKLinf}
\begin{split}
\Gamma^{(1)}_{inf} = & M \cdot \underset{\theta \in (1-m\varepsilon, 1+ M\varepsilon)}{\inf}\left[ -\frac{1}{\theta^2}\log e\right]\\
 = & -M \cdot \underset{\theta \in (1-m\varepsilon, 1+ M\varepsilon)}{\sup} \frac{1}{\theta^2} \log e\\
= & -M \cdot \frac{\log e}{\underset{\theta \in (1-m\varepsilon, 1+ M\varepsilon)}{\inf} \theta^2}\\
= & \frac{-M \log e}{(1 - \varepsilon m)^2},
\end{split}
\end{equation}
\begin{equation}
\begin{split}
\Gamma^2_{sup} = & -m \cdot \underset{\theta \in (1-m\varepsilon, 1+ M\varepsilon)}{\inf}\left[ -\frac{2}{\theta^3}\log e\right]\\
= & m \cdot \underset{\theta \in (1-m\varepsilon, 1+ M\varepsilon)}{\sup} \frac{2}{\theta^3} \log e\\
= & m \frac{2\log e}{\underset{\theta \in (1-m\varepsilon, 1+ M\varepsilon)}{\inf} \theta^3}\\
= & \frac{2m \log e}{(1 - \varepsilon m)^3},
\end{split}
\end{equation}
\begin{equation}
\begin{split}
\Gamma^2_{inf}
= & M \cdot \underset{\theta \in (1-m\varepsilon, 1+ M\varepsilon)}{\inf}\left[ -\frac{2}{\theta^3}\log e\right]\\
= & - M \cdot \underset{\theta \in (1-m\varepsilon, 1+ M\varepsilon)}{\sup} \frac{2}{\theta^3} \log e\\
= & - M \frac{2\log e}{\underset{\theta \in (1-m\varepsilon, 1+ M\varepsilon)}{\inf} \theta^3}\\
= & \frac{-2M \log e}{(1 - \varepsilon m)^3}.
\end{split}
\end{equation}
Then, we further have
\begin{equation}\label{cinequality1}
c_1 \geq \frac{\varepsilon m }{3(1-\varepsilon m)^2}, c_2 \geq \frac{2\varepsilon m }{3(1-\varepsilon m)^3},
\end{equation}
and
\begin{equation}\label{cinequality2}
\bar{c}_1 \leq \frac{- \varepsilon M }{3(1-\varepsilon m)^2}, \bar{c}_2 \leq \frac{-2\varepsilon M}{3(1-\varepsilon m)^3}.
\end{equation}
The inequality from Theorem \ref{differentablef} for the pair of $D(P_1\|P_0)$ and $D(P_0\|P_1)$ is expressed as
\begin{equation}\label{g1g2inequality1}
\frac{1+\bar{c}_1}{1+c_2} \leq \frac{D(P_1\| P_0)}{D(P_0\|P_1)}
\leq \frac{1+c_1}{1+\bar{c}_2}.
\end{equation}
From the above values of $c_i, \bar{c}_i, i= 1,2$, we have the tightest bound for (\ref{g1g2inequality1}) by taking equalities for the inequalities of
 $c_1, \bar{c}_1, c_2, \bar{c}_2$ in (\ref{cinequality1}) and (\ref{cinequality2}):
 \begin{equation}\label{lowerbound11}
 \begin{split}
 \frac{D(P_1\| P_0)}{D(P_0\|P_1)} \geq \frac{1 - \frac{\varepsilon M}{3(1-\varepsilon m)^2}}{1 + \frac{2\varepsilon m}{3(1-\varepsilon m)^3}} = 1 - \frac{2\varepsilon m + \varepsilon M (1-\varepsilon m)}{3(1-\varepsilon m)^3 +2\varepsilon m} \end{split}
 \end{equation}
 \begin{equation}\label{upbound11}
 \begin{split}
 \frac{D(P_1\| P_0)}{D(P_0\|P_1)}
 \leq  & \frac{1 + \frac{\varepsilon m}{3(1-\varepsilon m)^2}}{1- \frac{2\varepsilon M}{3(1-\varepsilon m)^3}}
=  1 + \frac{\varepsilon m(1-\varepsilon m) + 2\varepsilon M}{3(1-\varepsilon m)^3 -2\varepsilon M}
 \end{split}
 \end{equation}

In Theorem 6 in \cite{Verdu2}, the authors use the ratio of $g_i(t) - g'_i(1)(t-1), i= 1, 2$ and the domain of $\frac{dP_1}{dP_0}$ to bound $\frac{D(P_1\|P_0)}{\bar{D}(P_0\|P_1)}$. Here, we explicitly bound it using the domain of $\frac{dP_1}{dP_0}$ and the quantity relationships between the second and third order derivatives of the $g_i,i = 1, 2$.
\begin{Remark}
Theorem 6 in \cite{Verdu2} provides the following bounds of $\frac{D(P_1\|P_0)}{D(P_0\|P_1)}$:
\begin{equation}\label{globalbounds2}
\kappa(\beta_2) \leq \frac{D(P_1\|P_0)}{D(P_0\|P_1)} \leq \kappa(\beta_1^{-1})
\end{equation}
where
\begin{equation}\label{globalbound3}
\kappa(t) = \frac{t\log t + (1-t)\log e}{(t-1)\log e -\log t}.
\end{equation}
 In the generalized $\varepsilon_{\gamma}$-neighborhood, we have $\beta_1 = \frac{1}{1+ \varepsilon M}$ and $\beta_2 = 1 - \varepsilon m$. From (\ref{globalbound3}), the numerator and denominator have the form as $g(t) - g'(1)(t-1)$. The tightest bounds can be evaluated at $\beta_2$ and $\beta_1^{-1}$ due to the monotonicity of $\kappa$ from Remark 11 in \cite{Verdu2}. Since $c_i,\bar{c}_i, i = 1,2$ are obtained by evaluating the truncated Taylor's formula of $g_1,g_2$, our bounds are definitely looser than (\ref{globalbounds2}). However, the inequalities of (\ref{globalbounds2}) rely on the monotonicity of $\kappa$ function in (\ref{globalbound3}). When $\kappa$ function involved with pair of $f$-divergence is not monotonic, our bounds in Theorem \ref{differentablef} are more convenient to calculate.
\end{Remark}
\subsubsection{$\chi^2(P_1\|P_0)$ and $D(P_1\|P_0)$}
Let $g_1(t) =  t^2 - 1$ and $g_2(t) =  t\log t$, we have $g''_2(t) = \frac{\log e}{t}$, $g''_1(t) = 2$, $g'''_2(t) = -\frac{1}{t^2}\log e$ and $g'''_1(t) = 0 $. For $\varepsilon \gamma \leq 1$, we have $g'''_1(\theta) < 0$ and $g'''_2(\theta) < 0$ for $\theta \in (1-\varepsilon m, 1+ \varepsilon M)$. From (\ref{Gammaeq0}) (\ref{GammaKLsup}) and (\ref{GammaKLinf}), we get
$\Gamma^{(1)}_{sup} = \Gamma^{(1)}_{inf} = 0$, $\Gamma^{(2)}_{sup} = \frac{m \log e}{(1 - \varepsilon m)^2}$, and $\Gamma^{(2)}_{inf} = \frac{-M \log e}{(1 - \varepsilon m)^2}$.
Thus, we have
\begin{equation}
c_1 \geq 0, \bar{c}_1 \leq 0
\end{equation}
and
\begin{equation}\label{cinequality3}
c_2 \geq  \frac{\varepsilon m \log e}{6(1-\varepsilon m)^2}, \bar{c}_2 \leq - \frac{\varepsilon M \log e}{6(1-\varepsilon m)^2}.
\end{equation}
Finally, let $c_1 = \bar{c}_1 = 0$, the inequality from Theorem \ref{differentablef} for the pair of $\chi^2(P_1\|P_0)$ and $D(P_1\|P_0)$ is
\begin{equation}\label{ineq2}
\frac{2\ln 2}{1+c_2} \leq \frac{\chi^2(P_1\| P_0)}{D(P_1\|P_0)}
\leq \frac{2\ln 2}{1+\bar{c}_2},
\end{equation}
and the tightest bound is obtained for (\ref{ineq2}) by taking equalities for the inequalities of
 $c_2, \bar{c}_2$ in (\ref{cinequality3}).
\subsubsection{$H_{\alpha}(P_1,P_0)$ and $D(P_1\|P_0)$}
 Let $g_1(t) = \frac{t^{\alpha} -1}{\alpha  - 1}$ and $g_2(t) =  t\log t$, then $g_1''(t) = \alpha t^{\alpha - 2}$, $g_1'''(t) = \alpha (\alpha -2)t^{\alpha - 3}$, $g''_2(t) = \frac{\log e}{t}$ and $g'''_2(t) = -\frac{1}{t^2}\log e$. For $\varepsilon \gamma \leq 1$, we have $g'''_2(\theta) < 0$ for $\theta \in (1-\varepsilon m, 1+ \varepsilon M)$, and the sign of $g'''_1(\theta)$ depends on $\alpha$. From (\ref{Gammasup1}) and (\ref{Gammasup2}), we can get $\Gamma_{sup}^{(1)}$ as the same as (\ref{gammasupHellinger}), and $\Gamma_{inf}^{(1)}$ is calculated as
 \begin{equation}
 \begin{split}
   \Gamma_{inf}^{(1)} = &\underset{ \underset{\theta \in (1-m\varepsilon, 1+ M\varepsilon)}{ \eta \in (-m, M)}}{\inf} g'''_1(\theta)\cdot\eta\\
    = &  \underset{ \underset{\theta \in (1-\varepsilon m,1+ \varepsilon M)}{ \eta \in (-m, M)}}{\inf} \eta\alpha (\alpha -2)\theta^{\alpha - 3}
 \end{split}
 \end{equation}
 \begin{equation}
 = \left\{
 \begin{aligned}
        &M \times \underset{\theta \in (1-\varepsilon m,1+ \varepsilon M)}{\inf} \alpha (\alpha -2)\theta^{\alpha - 3}, \\
           &\,  \  \  \  \  \  \  \  \  \  \  \  \  \  \  \  \  \  \  \  \  \  \  \  \  \  \  \  \  \  \  \  \  \  \  \  \  \  \, 0 < \alpha < 2, \alpha \neq 1; \\
               &0 \,  \  \  \  \  \  \  \  \  \  \  \  \  \  \  \  \  \  \  \  \  \  \  \  \  \  \  \  \  \  \  \  \  \  \  \  \  \  \  \  \  \  \  \  \  \  \  \  \  \  \,  \alpha = 2; \\
      &-m \times \underset{\theta \in (1-\varepsilon m,1+ \varepsilon M)}{\sup} \alpha (\alpha -2)\theta^{\alpha - 3}, \, \ \,2 < \alpha < 3;\\
      &-m \times \underset{\theta \in (1-\varepsilon m,1+ \varepsilon M)}{\sup} \alpha (\alpha -2)\theta^{\alpha - 3},   \, \ \  \  \  \  \, \alpha \geq 3 \\
 \end{aligned}
    \right.
    \end{equation}
    \begin{equation}
 = \left\{
 \begin{aligned}
        &M \alpha (\alpha -2)\times \underset{\theta \in (1-\varepsilon m,1+ \varepsilon M)}{\inf} \frac{1}{\theta^{3-\alpha }}, \\
           &\,  \  \  \  \  \  \  \  \  \  \  \  \  \  \  \  \  \  \  \  \  \  \  \  \  \  \  \  \  \  \  \  \  \  \  \  \  \  \  \, 0 < \alpha < 2, \alpha \neq 1; \\
               &0 \,  \  \  \  \  \  \  \  \  \  \  \  \  \  \  \  \  \  \  \  \  \  \  \  \  \  \  \  \  \  \  \  \  \  \  \  \  \  \  \  \  \  \  \  \  \  \  \  \  \  \  \,  \alpha = 2; \\
      &-m  \alpha (\alpha -2)\times \underset{\theta \in (1-\varepsilon m,1+ \varepsilon M)}{\sup}\frac{1}{\theta^{3-\alpha }}, \, \  \,2 < \alpha < 3;\\
      &-m\alpha (\alpha -2) \times \underset{\theta \in (1-\varepsilon m,1+ \varepsilon M)}{\sup} \theta^{\alpha - 3},   \, \ \  \  \  \  \  \  \, \alpha \geq 3 \\
 \end{aligned}
    \right.
    \end{equation}
\begin{equation}
\begin{split}
= &  \begin{cases}
        M \cdot \frac{\alpha(\alpha -2)}{(1 + \varepsilon M)^{3-\alpha}}, & 0 < \alpha < 2, \alpha \neq 1; \\
       0, & \alpha  = 2; \\
     -m \cdot \frac{\alpha(\alpha -2)}{(1 - \varepsilon m)^{3-\alpha}}, & 2 < \alpha < 3;\\
     -m \cdot \alpha(\alpha -2)(1 + \varepsilon M)^{\alpha- 3}, &  \alpha \geq 3.\\
      \end{cases}
   \end{split}
 \end{equation}
The quantities of $\Gamma_{sup}^{(2)}$ and $\Gamma_{inf}^{(2)}$ are the same as (\ref{GammaKLsup}) and (\ref{GammaKLinf}), respectively. Therefore, the inequality of $c_1$ is the same as (\ref{cVTHalpha}), and we further have
\begin{equation}
\begin{split}
\bar{c}_1\leq \frac{\varepsilon  \Gamma^{(1)}_{inf}}{3g_1''(1)}
= & \frac{\varepsilon \Gamma^{(1)}_{inf}}{3 \alpha}\\
= &  \begin{cases}
     \frac{\varepsilon M(\alpha -2)}{3(1 + \varepsilon M)^{3-\alpha}}, & 0 < \alpha < 2, \alpha \neq 1; \\
       0, & \alpha  = 2; \\
     -\frac{\varepsilon m(\alpha -2)}{3(1 - \varepsilon m)^{3-\alpha}}, & 2 < \alpha < 3;\\
 -\frac{\varepsilon m(\alpha -2)(1 + \varepsilon M)^{\alpha- 3}}{3}, &  \alpha \geq 3.\\
      \end{cases}
\end{split}
\end{equation}
and
\begin{equation}
c_2 \geq \frac{\varepsilon m }{3(1-\varepsilon m)^2}, \bar{c}_2 \leq \frac{- \varepsilon M }{3(1-\varepsilon m)^2}.
\end{equation}
Finally, the inequality from Theorem \ref{differentablef} for the pair of $H_{\alpha}(P_1,P_0)$ and $D(P_1\|P_0)$ is
\begin{equation}\label{ineq2}
\alpha \ln 2 \cdot \frac{(1+\bar{c}_1)}{(1+c_2)}\leq \frac{H_{\alpha}(P_1\| P_0)}{D(P_1\|P_0)}
\leq \alpha \ln 2 \cdot \frac{(1+c_1)}{(1+\bar{c}_2)},
\end{equation}

\subsection{An Example}\label{Examplesection}

In the following, the equivalent conditions with the assumption in Corollary \ref{localcondition} are illustrated by an example, we will show that the bound in  Corollary \ref{localcondition} is more tight than the existing ones in literature. 

There are $n$ elements $x_1, x_2, \cdots, x_n$ in $\mathcal{X}$, and the probability mass of $P_1$ and $P_0$ satisfy
\begin{equation}
P_1(x_i) = p_i, i = 1,\cdots,n.
\end{equation}
\begin{equation}
P_0(x_i) = q_i, i = 1,\cdots,n.
\end{equation}
With given $\varepsilon$, the function $h$ is expressed as
\begin{equation}
h(x_i) = \frac{p_i - q_i}{\varepsilon \cdot q_i},
\end{equation}
With given $(m,M)$, then the requirement on $h$ becomes
\begin{equation}
-m \leq h(x_i) \leq M
\end{equation}
which is equivalent to
\begin{equation}\label{im1}
1 - m \cdot \varepsilon \leq \frac{p_i}{q_i} \leq 1 + M\cdot \varepsilon, i = 1,2,\cdots n.
\end{equation}
The generalized quasi-$\varepsilon$-neighborhood is closely related to the strongly $\delta$-typical set $\mathcal{T}_{\delta}(P_0)$, which is given by
\begin{equation}
\mathcal{T}_{\delta}(P_0) = \left\{P_1: \forall x \in \mathcal{X}, |P_1(x) - P_0(x)| \leq \delta P_0(x)\right\}.
\end{equation}
It is obvious that $\delta = \varepsilon \gamma$ in our framework.
Consider $\Delta$ in this case. It is rewritten as
\begin{equation}
\begin{split}
 \Delta = \sum_{i=1}^n q_i|h(x_i)|^2
 =  \frac{1}{\varepsilon^2} \sum_i^n\frac{(p_i - q_i)^2}{q_i}.
 \end{split}
 \end{equation}
 From (\ref{im1}), $-m \varepsilon q_i \leq p_i - q_i \leq M\varepsilon q_i$ holds for all $i$, we further have
 \begin{equation}
 \Delta \leq \frac{1}{\varepsilon^2} \sum_i^n \frac{\varepsilon^2 \gamma^2 q_i^2}{q_i} \leq \gamma^2.
 \end{equation}
\textbf{Exmaple.} Let $P_1$ and $P_0$ be both generalized Bernoulli distributions (or categorical distributions) whose probability mass functions are
\begin{equation}
P_0(x = i) = \frac{1}{n}, \, \  \  \  \, i = 1,\cdots,n,
\end{equation}
and
\begin{equation}
\begin{split}
&P_1(x = 1) = \frac{1}{n} +  \frac{1}{mn}, \, \  \  \  \, P_1(x  = 2) =  \frac{1}{n} - \frac{1}{mn},\\
&P_1(x = i) = \frac{1}{n}, \, \  \  \  \, i = 3,\cdots,n,
\end{split}
\end{equation}
where $n > 2$ and $m > \max\{10, n\}$ is a large integer . Then $TV(P_1,P_0)$ and $D(P_1\|P_0)$ between  $P_1$ and $P_0$ are
\begin{equation}
TV(P_1,P_0) = \frac{1}{mn},
\end{equation}
\begin{equation}
\begin{split}
D(P_1\|P_0) = &\frac{1}{n}(1 + \frac{1}{m}) \log (1 + \frac{1}{m})\\
 + & \frac{1}{n}(1 - \frac{1}{m}) \log (1 - \frac{1}{m})\\
= &\frac{1}{n}(1 + \frac{1}{m}) (\frac{1}{m} - \frac{1}{2* m^2} + \frac{1}{3*m^3} - \cdots)\log e \\
+ & \frac{1}{n}(1 - \frac{1}{m})(-\frac{1}{m} - \frac{1}{2* m^2} - \frac{1}{3*m^3} - \cdots)\log e\\
\approx & \frac{1}{n}(\frac{1}{m^2} + \frac{2}{3m^4})\log e.
\end{split}
\end{equation}
Let $\varepsilon = \frac{1}{k}$ where $ 10 < k < m$ and $\gamma = 1$, then we have
\begin{equation}
\begin{split}
h(1) = \frac{\frac{1}{mn}}{\frac{1}{n}\cdot \frac{1}{k}} = \frac{k}{m}, \, \  \  \, h(2) = -\frac{k}{m},\\
 \, \  \  \, h(i) = 0, \, \ \, i = 3,\cdots,n.
\end{split}
\end{equation}
and
\begin{equation}
\Delta = \frac{2k^2}{nm^2},\, \  \  \, \underset{i}{\min} P_0(i) = \frac{1}{n}.
\end{equation}
Finally,
\begin{equation}
\begin{split}
&\frac{1}{2}\sqrt{D(P_1\| P_0)\Delta \ln 2}\\
\approx & \frac{1}{2} \sqrt{\frac{1}{n}(\frac{1}{m^2} + \frac{2}{3m^4}) \frac{2k^2}{nm^2} \cdot \log e \cdot \ln 2} \\
\leq  &\frac{\sqrt{3}}{2}  \times \frac{k}{n m^2} \\
\leq &\frac{1}{mn} = TV(P_1, P_0)
\end{split}
\end{equation}
Now we compare the upper bounds of $D(P_1\|P_0)$ from (\ref{lowerboundKL1}) and (\ref{finitealphabet}) when $m$ is sufficiently small in above case. The upper bound of $D(P_1\|P_0)$ from (\ref{lowerboundKL1}) is expressed as
\begin{equation}\label{KLupper1}
\begin{split}
& \frac{2(1+c)\gamma^2 TV^2(P_1,P_0)\log e}{\Delta}\\
= & \frac{2(1+c)}{m^2n^2}\frac{nm^2}{2k^2} \cdot \log e\\
= &\frac{1+c}{nk^2}\cdot \log e
\end{split}
\end{equation}
From the inequality $\ln (1+x) \geq \frac{x}{1+x}, \, \  \,x > 0$, the upper bound from (\ref{finitealphabet}) satisfies
\begin{equation}\label{KLupper2}
\begin{split}
&\log \left(1 + \frac{2TV^2(P_1,P_0)}{\min P_0(i)}\right) \\
= & \log (1 + 2n V^2_T(P_1,P_0))\\
= &  \log \left(1 + \frac{2n}{m^2n^2}\right) \\
\geq & \frac{\frac{2}{m^2n}}{1 + \frac{2}{m^2n}} \cdot \log e\\
= & \frac{2}{2 + m^2n}\cdot \log e
\end{split}
\end{equation}
If \begin{equation}\label{conditiontighter}
2k^2 >  (1+c)m^2 + \frac{2(1+c)}{n},
\end{equation} we further have
\begin{equation}
\frac{1+c}{nk^2}\cdot \log e \leq \frac{2}{2 + m^2n}\cdot \log e
\end{equation}

Thus, the upper bound from (\ref{lowerboundKL1}) is a tighter upper bound than that from (\ref{finitealphabet}) in this example.

\section{Applications in Asymptotic Distributions of $f$-Divergence Test of Goodness of Fit}
In this section, we apply the inequalities to extend the existing asymptotic distributions of a family of statistics used in testing the goodness of fit \cite{Vajda0}. The extensions are from two aspects. First, different from the assumption that the underlying "cell distribution" \cite{Serfling} is uniform, our conclusions are suitable for any cell distribution. Second, we present the asymptotic of TV statistics based goodness of fit, while previous $f$-divergence statistics have common characteristics as being second order differentiable.

Consider a sequence of $n$ independent trails, with $k$ possible outcomes for each trail. These possible outcomes are from a partition of $\mathcal{R}^m$ as $\{B_1, B_2, \cdots, B_k\}$, and the probability that a given observation $X_i$ lies in $B_j$ is $p_j = P(X_i^{-1}(B_j))$. Denote $p = (p_1, \cdots, p_k)$ be the cell distribution. With the $n$ trails, the relative frequency vector $ \hat{p}_n = (\frac{n_1}{n},\cdots, \frac{n_k}{n})$ is a multinomial distribution $(n; p_1, \cdots, p_k)$. It is well known \cite{Serfling} that the chi-squared statistic
\begin{equation}
T_n = \sum_{i=1}^{k} \frac{(n_i - np_i)^2}{np_i} = n \chi^2(p_n,p)\stackrel{d}{\longrightarrow}  \chi^2_{k-1} \, \  \, \text{as} \, \  \  \, n \rightarrow \infty.
\end{equation}
\begin{Corollary}
For any discrete distribution $p$ and its empirical distribution $\hat{p}_n$ defined above, we have for these $D_g$ in the second type (that has third order derivative around $t = 1$),
\begin{equation}
\frac{2n}{g''(1)} D_g (\hat{p}_n, p) \stackrel{d}{\longrightarrow}  \chi^2_{k-1}.
\end{equation}
where $\stackrel{d}{\longrightarrow}$ denotes convergence in distribution.
\end{Corollary}
\begin{proof}
From law of large numbers, we have $\hat{p}_n \rightarrow p, \,  \  \, a.s$ as $n \rightarrow \infty$ and it leads to
\begin{equation}
\hat{p}_n(k) \rightarrow p_k \, \ \  \, \forall k,
\end{equation}
which is equivalent to
\begin{equation}\label{disc1}
\frac{\hat{p}_n(k)}{p_k} - 1 \rightarrow 0 \, \ \  \, \forall k
\end{equation}
for fixed $p$.
Let $\hat{p}$ and $p$ be the corresponding $P_1$ and $P_0$ in the generalized quasi-$\varepsilon$-neighborhood. Then (\ref{disc1}) further implies that $\gamma \rightarrow 0$ for any given $\varepsilon$.
From Theorem \ref{differentablef}, let $D_{g_2}$ be $\chi^2$-Divergence, we have
\begin{equation}
\frac{1+\bar{c}_1}{2(1+c_2)}g''_1 \leq \frac{D_{g_1}(P_1,P_0)}{\chi^2(P_1,P_0)} \leq \frac{1+c_1}{2(1+\bar{c}_2)} g''_1,
\end{equation} Therefore, from Remark \ref{Remark4} we have
\begin{equation}
\frac{D_{g_1}(\hat{p}_n, p)}{\chi^2(\hat{p}_n,p)} \rightarrow \frac{g''(1)}{2}.
\end{equation}
Consequently,
\begin{equation}
\frac{2n}{g''(1)} D_g (\hat{p}_n, p) \stackrel{d}{\longrightarrow}  \chi^2_{k-1}.
\end{equation}
\end{proof}
The above conclusion is a generalization of Corollary 3 in \cite{Vajda} to allow $p$ to be any discrete distribution instead of uniform distribution.
It is also known that for any distribution pair $P$ and $Q$ \cite{Verdu2}\cite{Reid},
\begin{equation}\label{lower1}
TV^2(P,Q) \leq \frac{1}{4}\chi^2(P,Q), \, \  \  \  \  \, \text{if} \, \  \  \, TV(P,Q) \leq \frac{1}{2}.
\end{equation}
From (\ref{chi2VT}), we know in the generalized quasi-$\varepsilon$-neighborhood, we have
\begin{equation}\label{upper1}
\frac{1}{4}\frac{\Delta \chi^2(P_1\|P_0)}{\gamma^2} \leq TV^2(P_1,P_0).
\end{equation}
Applying the above bounds for the distribution pair of $\hat{p}_n$ and $p$, we arrive at the following asymptotic distribution of TV version of goodness of fit.
\begin{Corollary}
there exists some $c_t$ such that $1  \leq c_t \leq \frac{\gamma^2}{\Delta}$,
\begin{equation}
4nc_t TV^2(p_n,p) \stackrel{d}{\longrightarrow}  \chi^2_{k-1} \, \  \, \text{as} \, \  \  \, n \rightarrow \infty.
\end{equation}
\end{Corollary}
Following Theorem 1 in \cite{Tumanyan}, we further have the following corollaries on asymptotic distributions of $f$-divergence test of goodness of fit when the number of observations and the number of cells increase simultaneously. 
\begin{Corollary}
For any discrete distribution $p$ and its empirical distribution $\hat{p}_n$ defined above, if $n$ and $k$ increase simultaneously so that
\begin{equation}
\underset{1\leq i \leq k}{\min} np_i\rightarrow \infty,
\end{equation}
we have for these $D_g$ in the second type,
\begin{equation}
\frac{2n D_g(p_n,p)/g''(1) - (k-1)}{\sqrt{2(k-1)}} \stackrel{d}{\longrightarrow}  N(0,1), \, \  \  \, n \rightarrow \infty.
\end{equation}
\end{Corollary}
\begin{Corollary}
there exists some $c_t$ such that $1  \leq c_t \leq \frac{\gamma^2}{\Delta}$,
\begin{equation}
 \frac{4nc_t TV^2(p_n,p) - (k-1)}{\sqrt{2(k-1)}} \stackrel{d}{\longrightarrow}  N(0,1), \, \  \  \, n \rightarrow \infty.
\end{equation}
\end{Corollary}

\section{Conclusion}
This study has established a systematic framework for bounding f-divergences in local distributional regimes through three key advancements. The proposed generalized quasi-$\varepsilon_{(M,m)}$-neighborhood extends existing proximity models by incorporating parametric flexibility $(M,m)$, enabling unified treatment of discrete/continuous distributions where $dP_1/dP_0 \approx 1$.
By classifying $f$-divergences according to first-order differentiability at unity, we derived Taylor-based inequalities that subsume classical $\chi^2$-divergence bounds as special cases.
The reverse Pinsker-type inequalities demonstrate particular efficacy in goodness-of-fit testing, offering computable bounds for asymptotic analysis when testing hypotheses with proximate alternatives. 
Future directions include extending this framework to non-differentiable f-divergences and exploring applications in differential privacy bounds. The methodology's parametric adaptability suggests promising extensions to high-dimensional statistical learning problems.


\begin{thebibliography}{99}
\bibitem{Csiszar0} I.\, Csisz\'{a}r, ``Information mesaures: A Critical Survey, " in \emph{Trans. 7th Prague Conference on Inf. Theory,} Academia: Prague, 1977; pp. 73–86.
\bibitem{Csiszar1} I.\, Csisz\'{a}r and P.Shields, ``Information theory and statistics: A tutorial," \emph{Found. Trends Commun. Inf. Theory,} vol. 1, no. .4, pp. 417-528, 2004. 
\bibitem{Verdu} S.\.Verd\'{u}, ``Total variation distance and the distribution of relative information,” in \emph{2014 Information Theory and Applications Workshop} (ITA), 2014, pp. 1-3.
\bibitem{Verdu2} I.\,Sason and S.\,Verd\'{u},``$f$-Divergence Inequalities," \emph{IEEE Trans. Inf. Theory}, Vol.\,62, No.\,11, pp.\,5973-6006, Nov.\,2016.
\bibitem{Sreekumar} S.\,Sreekumar, Z.\,Goldfeld and K.\,Kato,``Limit Distribution Theory for $f$-Divergences," \emph{IEEE Trans. Inf. Theory}, Vol.\,70, No.\,2, pp.\,1233-1267, Feb.\,2024.
\bibitem{Guntuboyina} A.\,Guntuboyina, ``Lower bounds for the minimax risk using $f$-divergences, and applications,"  \emph{IEEE Trans. Inf. Theory}, Vol.\,57, No.\,1, pp.\,2386-2399, Feb.\,2011.
\bibitem{Guntuboyina2} A.\, Guntuboyina, ``Minimax lower bounds," Ph.D. dissertation, Dept. Statist., Yale Univ., New Haven, CT, USA, 2011. 
\bibitem{Reid} M. D. Reid and R. C. Williamson, ``Generalized Pinsker inequalities,"  in \emph{ Proc. 22nd Annu. Conf. Learn. Theory }, 2009, pp. 1-10.
\bibitem{Reid2} M. D. Reid and R. C. Williamson, ``Information, divergence and risk for binary experiments," \emph{Journal of Machine Learning Research}, vol. 12, no. 3, pp. 731–817, Mar. 2011.

\bibitem{Csiszar2} I.\, Csisz\'{a}r, ``A Note on Jensen's inequality," \emph{Studia Scientarium Math. Hungarica,} vol. 1, pp. 185-188, Jan. 1966. 
\bibitem{Vajda4} I. Vajda, ``Note on discrimination information and variation," \emph{IEEE Trans. Inf. Theory}, Vol.\,16, No.\,6, pp.\,771-773, Sep.\,1970.
\bibitem{Gilardoni}G. L. Gilardoni, `` On the minimum $f$-divergence for given total vaiation," \emph{Comptes Rendus Math.,} vol. 343, issue. 11-12, pp. 763-766, 2006. 
\bibitem{Topsphie} F.\,Tops$\phi$e, ``Some inequalities for information divergence and related measures of discrimination,"  \emph{IEEE Trans. Inf. Theory}, Vol.\,46, No.\,4, pp.\,1602-1609, Jul.\,2009.
\bibitem{Fedotov} A. Fedotov, P.\,Harremoës, and F. Top$\phi$e, ``Refinements of Pinsker's inequality, "\emph{IEEE Trans. Inf. Theory}, Vol.\,49, No.\,6, pp.\,1491-1498, Jun.\,2003.
\bibitem{Dragomir} S.\,S.\,Dargomir, ``An upper bound for the Csisz\'{a}r $f$-divergence in terms of the variational distance and applications," in Inequalities for Csisz\'{a}r $f$-divergence in Information Theory. RGMIA Monographs, S.\,S.\,Dargomir and T.\,M.\,Rassias, Eds. Melbourne VIC. Australia: Victoria University, Australia, 2000.
\bibitem{Vajda1} P.\,Harremoës and I.\,Vajda,  ``On Pairs of $f$-Divergences And Their Joint Range," \emph{IEEE Trans. Inf. Theory}, Vol.\,57, No.\,6, pp.\,3230-3235, Jun.\,2011.
\bibitem{Binette} O.\, Binette, “A Note on Reverse Pinsker Inequalities,” \emph{IEEE Trans. Inf. Theory}, vol. 65, no. 7, pp. 4094-4096, 2019.
\bibitem{Raginsky} M.\, Raginsky, ``Strong Data Processing Inequalities and $\Phi$-Sobolev Inequalities for Discrete Channels," \emph{IEEE Trans. Inf. Theory}, vol. 62, no. 6, pp. 3355-3389, Jun. 2016.
\bibitem{Barnett} N.\,S.\,Barnett, P.\,Cerone, S.\,S.\,Dragomir and A.\,Sofo, ``Approximating Csisz\'{a}r $f$-divergence by the use of Taylor's Formula with integral reminder," \emph{Mathematical Inequalities and Applications}, 5:417-434,2002. 
\bibitem{Nielsen0} F.\, Nielsen and R.\, Nock, ``On the Chi Square and Higher-Order Chi Distances for Approximating f-Divergences," \emph{IEEE Signal Processing Letters}, Vol.21, No. 1, Jan. 2014.  
\bibitem{Sason} I.\,Sason, ``On Reverse Pinsker Inequalities," ArXiv eprints, 2015, arXiv: 1503.07118.
\bibitem{Gyorfi} L.\,Gyorfi, G.\,Morvai, and I.\,Vajda,  ``Information-theoretic methods in testing the goodness of fit," in \emph{IEEE International Symposium on Inforamtion Theory,} Sorrento, Italy, Jun, 2000. 
\bibitem{Bor99} A. A. Borovkov, {\sl Mathematical Statistics}. CRC Press, 1999.
\bibitem{Vajda0} M.\,C.\,Pardo and I.\,Vajda,  ``On Asymptotic Properties of Information-Theoretic Divergences," \emph{IEEE Trans. Inf. Theory}, Vol.\,49, No.\,7, pp.\,1860-1868, Jul.\,2003.
\bibitem{Yury1} Y. Polyanskiy, "Information theory methods in statistics and computer science," MIT course page, 2019-2020.
\bibitem{Yury4} Y.\,Polyanskiy, Y. Wu, {\sl Information Theory: From Coding to Learning}. Cambridge University Press, 2024.
\bibitem{George1} I.\,George, A.\,Zheng and A.\,Bansal, ``Divergence Inequalities from Multivariate Taylor's Theorem, "  \emph{2024  IEEE Information Theory Workshop} (ITW), 2024, pp. 561-566.
\bibitem{George2} I.\,George, A.\,Zheng and A.\,Bansal, ``Divergence inequalities with applications in ergodic theory," 2024. https:// arxiv.org/abs/2411.17241.
\bibitem{Huang} S. Huang, C. Suh, and L. Zheng, ``Euclidean information theory of networks,” \emph{IEEE Trans. Inf. Theory}, vol. 61, no. 12, pp. 6795–6814, 2015.
\bibitem{Sason1}I.\,Sason, ``On f-Divergences: Integral Representations, Local Behavior, and Inequalities," \emph{Entropy}, 2018, vol. 20, issue 5, 383.
\bibitem{Nielsen} F.Nielsen,``Statistical Divergences between Densities of Truncated Exponential Families with Nested Supports: Duo Bregman and Duo Jensen Divergences," \emph{Entropy}, 2022, vol. 24, 421. 
\bibitem{Hewitt} H.\,Hewitt and K.\,Stromberg, {\sl Real and Abstract Analysis}. Berlin, Germany: Springer, 1965.
\bibitem{Rockafellar} R.\,Tyrrell Rockafellar, {\sl Convex Analysis}. Princeton University Press, 1997.
\bibitem{Liese} F.\,Liese and I.\, Vajda, ``On Divergences and Informations in Statistics and Information Theory," \emph{IEEE Trans. Inf. Theory}, Vol.\,52, No.\,10, pp.\,4394-4412, Otc.\,2006.
\bibitem{Basu} A.\,Basu, H.\,Shioya, and C.\,Park, {\sl Statistical Inference: The Minimum Distance Approach} (Chapman $\&$ Hall/CRC Monographs on Statistics $\&$ Applied Probability). vol. 120, Boca Raton, FL, USA: CRC Press, Jun. 2011.
\bibitem{Yu1} X. Yu, S. Wei and Y. Luo, ``Finite Blocklength Analysis of Gaussian Random Coding in AWGN Channels under Covert Constraint," \emph{IEEE Transactions on Information Forensics and Security}, Vol. 16, pp.\, 1261-1274, 2021.
\bibitem{Hamkins} J. Hamkins and K. Zeger, ``Gaussian Source Coding with Spherical Codes,"\emph{IEEE Trans. Inf. Theory}, Vol.\,48, No.\,11, pp.\,2980-2989, Nov. 2002.
\bibitem{Vajda} I.\,Vajda,{\sl Convex Statistical Distances}. Leipzig, Germany, Springer, 1987.
\bibitem{Serfling} R.\,J.\,Serfling, {\sl Approximation Theorems of Mathematical Statistics}. John Wiley\&Sons, 1980.

\bibitem{Tumanyan} S. H. Tumanyan, ``Asymptotic distribution of the $\chi^2$ criterion when the number of observations and number of groups increase simultaneously," \emph{Theory of Probability and Its Applications}, vol. 1, 1956.
\bibitem{Jurg} J.\, H\"{u}sler, R.\,Y.\,Liu and K.\,Singh, ``A Formula for the Tail Probability of a Multivariate Normal Distribution and Its Applications," \emph{Journal of Multivariate Analysis}, Vol.\,82, pp.\,422-430,  2002.




\end{thebibliography}
\end{document}